\documentclass[numberedappendix,iop]{emulateapj}
\usepackage[varg]{txfonts}
\newcommand{\se}[1]{Section\ \ref{sec:#1}}

\newcommand{\Se}[1]{Section\ \ref{sec:#1}}

\newcommand{\eq}[1]{Equation\ (\ref{eq:#1})}
\newcommand{\eqp}[1]{Equation\ (\ref{eq:#1})}

\newcommand{\Eq}[1]{Equation\ (\ref{eq:#1})}

\newcommand{\fg}[1]{Figure\ \ref{fig:#1}}

\newcommand{\Fg}[1]{Figure\ \ref{fig:#1}}
\newcommand{\Tb}[1]{\mbox{Table\ \ref{tab:#1}}}

\newcommand{\ie}{i.e.,}
\newcommand{\eg}{e.g.,}
\newcommand{\etc}{etc}
\newcommand{\cf}{cf.}
\newcommand{\micr}{\ensuremath{\mu\mathrm{m}}}

\shorttitle{The fate of planetesimals in turbulent disks. II}
\shortauthors{Ormel \& Okuzumi}

\begin{document}


\title{The fate of planetesimals in turbulent disks with dead zones. II. \\ Limits on the viability of runaway accretion}


\author{C.W. Ormel\altaffilmark{1}}
\affil{Astronomy Department, University of California, Berkeley, CA 94720}
\email{ormel@astro.berkeley.edu}

\author{S. Okuzumi\altaffilmark{2,3}}
\affil{Department of Earth and Planetary Sciences, Tokyo Institute of Technology, Meguro-ku, Tokyo, 152-8551} 
\email{okuzumi@geo.titech.ac.jp}


\altaffiltext{1}{Hubble Fellow}
\altaffiltext{2}{JSPS Superlative Research Fellow}
\altaffiltext{3}{Department of Physics, Nagoya University, Nagoya, Aichi 464-8602, Japan}

%Scattering of planetesimals causes a planet to migrate.  
\begin{abstract}
A critical phase in the standard model for planet formation is the runaway growth phase. During runaway growth bodies in the 0.1--100 km size range (planetesimals) quickly produce a number of much larger seeds. The runaway growth phase is essential for planet formation as the emergent planetary embryos can accrete the leftover planetesimals at large gravitational focusing factors.  However, torques resulting from turbulence-induced density fluctuations may violate the criterion for the onset of runaway growth, which is that the magnitude of the planetesimals' random (eccentric) motions are less than their escape velocity. This condition represents a more stringent constraint than the condition that planetesimals survive their mutual collisions.  To investigate the effects of MRI turbulence on the viability of the runaway growth scenario, we apply our semi-analytical recipes of Paper I, which we augment by a coagulation/fragmentation model for the dust component. We find that the surface area-equivalent abundance of $0.1$ \micr\ particles is reduced by factors $10^2$--$10^3$, which tends to render the dust irrelevant to the turbulence. We express the turbulent activity in the midplane regions in terms of a size $s_\mathrm{run}$ above which planetesimals will experience runaway growth. We find that $s_\mathrm{run}$ is mainly determined by the strength of the vertical net field that threads the disks and the disk radius. At disk radii beyond 5 AU, $s_\mathrm{run}$ becomes larger than $\sim$100 km and the collision times among these bodies longer than the duration of the nebula phase. Our findings imply that the classical, planetesimal-dominated, model for planet formation is not viable in the outer regions of a turbulent disk.

%Here we focus on the condition required for the onset of the runaway growth phase,   However, for a system to enjoy runaway growth, . 

%caused by MRI turbulence in the gas disk excites planetesimals, causing their destruction when the turbulent excitation is too large.
\end{abstract}





\section{Introduction}
\label{sec:intro}
Gas in protoplanetary disks is thought to be turbulent. Direct observational support for the turbulent nature of disks is difficult to gather as detecting subsonic turbulence is challenging (but see \citealt{HughesEtal2011,GuilloteauEtal2012} for recent, positive detections). The prime observational reason hinting a turbulent nature is that young, T-Tauri stars are active accretors ($\sim$$10^{-8}\ M_\sun\ \mathrm{yr}^{-1}$).  The molecular viscosity, $\nu_\mathrm{mol} \sim c_s \ell_\mathrm{mfp}$ where $c_s$ is the sound speed and $\ell_\mathrm{mfp}$ the mean-free-path the gas, is however too small to account for these large-scale transport phenomena. The magneto-rotational instability (MRI; \citealt{BalbusHawley1991}) is commonly accepted as the most promising mechanism to drive the angular-momentum transport. 

A key requirement for the MRI to operate is that disks must be sufficiently ionized. Although the required ionization levels are only tiny, they might not be met in the very dense midplane regions of the disks \citep{Gammie1996}. The extent of this dead zone and the resulting properties of the turbulence depend on the large-scale magnetic field ($B_{z0}$), whose (uncertain) strength derives back to the molecular cloud from which the star formed and its long-term evolution in the disk. The turbulent properties also depend on the resistivity of the gas, which is determined by the gas' column density and dust properties \citep{SanoEtal2000,IlgnerNelson2006}. As the midplane regions are the sites where planet formation takes place, characterizing the turbulence in dead zones is of prime importance. 

One manifestation of turbulence is that the gas density distribution becomes clumpy. Although in subsonic turbulence the magnitude of these density fluctuations is small, $\delta\rho/\rho \ll 1$, the cumulative effect of the ensuing stochastic torques profoundly affects the orbital parameters of solid bodies, \eg\ semi-major axis or eccentricity. For planets, these turbulence-induced density fluctuations have been invoked as a new, `random' migration mechanism \citep{LaughlinEtal2004i,Nelson2005,OgiharaEtal2007}. Likewise, the density fluctuations excite the motions of smaller $\sim$km-size bodies (planetesimals), thought to be the building blocks of planets. It was realized that ideal MRI turbulence would most likely destroy planetesimals through collisions \citep{Nelson2005,IdaEtal2008,NelsonGressel2010}. The underlying reason is that planetesimals in the 100 m--10 km size range are at their minimum in the strength curve \citep[\eg][]{BenzAsphaug1999} -- \ie\ when two of them collide, a relatively low velocity suffices to destroy the bodies. To overcome this destructive collisional activity, and to salvage their role as planetary building blocks,  dead zones have been suggested as `safe havens' for planetesimals \citep{GresselEtal2011,GresselEtal2012}. 

The survivability question of planetesimals obviously is important; but in this paper we will address another weakness of the standard paradigm for planet formation, \ie\ the core accretion model \citep{Safronov1969,Mizuno1980,PollackEtal1996}.  A critical assumption of this model is that a population of planetesimals undergoes a runaway growth (RG) phase.  RG is triggered when the random velocity dispersion of the system, $\sigma(\Delta v)$, falls below their escape velocity $v_\mathrm{esc}$ of the bodies:
\begin{equation}
  v_\mathrm{esc}
  = \sqrt{\frac{2G_N m}{s}}
  = \sqrt{\frac{8\pi G_N \rho_\bullet}{3}} s,
\end{equation}
where $G_N$ is Newton's gravitational constant, $m$ the mass of the body, $s$ its radius (size), and $\rho_\bullet$ the internal density.  RG has several beneficial consequences for planet formation. Firstly, when the RG-condition becomes satisfied ($\Delta v < v_\mathrm{esc}$) gravitational deflection boosts the collision cross section by a factor $(v_\mathrm{esc}/\Delta v)^2$ -- the gravitational focusing factor -- and growth timescales are reduced accordingly \citep{WetherillStewart1989}. Furthermore, the gravitational focusing causes the biggest bodies to enjoy the largest growth rates, resulting in a quick formation of a few planetary embryos. Once initiated, RG is self-sustained: the bodies that enjoy large growth rates will continue to do so, because their $v_\mathrm{esc}$ increases with mass. At later times viscous stirring (by the embryos) will stabilize or decrease focusing factors, but these stay nonetheless much larger than unity \citep[\eg][]{KokuboIda2000,Chambers2008}. The outcome of runaway growth is a two component system where embryos sweep-up the leftover planetesimals at large focusing factors \citep{KokuboIda1998,OrmelEtal2010i}. The later formation phases are not without difficulties (to form giant planets either big cores or a very efficient cooling mechanism for the embryos atmospheres is required); but the two-component outcome is altogether beneficial for planet formation and a cornerstone of the core accretion paradigm.\footnote{Exception are dense, close in systems, where collision timescales are already short enough even without gravitational focusing \citep{ChiangLaughlin2012}.}

Driven by the idea that the final doubling of the solid core's mass is the bottleneck and therefore the more interesting area to pursue, many works just start from this setup \citep{ThommesEtal2003,Chambers2008,MordasiniEtal2009}. But the implicit assumption in these works is that an ensemble of 0.1--100 km-size planetesimals did enjoy a runaway growth phase. Therefore, as outlined above, the condition for runaway growth, $\Delta v < v_\mathrm{esc}$, must have been met at some earlier time; and for this we need the gaseous disk to be sufficiently quiescent. In this way, turbulence constrains planet formation models on a very fundamental level.

Therefore, an understanding of the dead zone physics is important. Previously \citet{OkuzumiHirose2011} (henceforth OH11) have conducted Ohmic-resistive MRI simulations, and constructed a toy model to quantify the turbulent activity throughout the vertical extent (active layers and dead zone). In \citet{OkuzumiOrmel2013} (henceforth, Paper I) we have extended these set of recipes to match the planetesimal excitation behavior seen in the simulation of \citet*{GresselEtal2012} (henceforth, GNT12). We achieved excellent agreement. This allows us to obtain the rate of planetesimal stirring by the turbulence \textit{a priori}, \ie\ without resorting to numerically expensive MRI simulations.
%In this paper, we will apply these results to
%Although the subsonic turbulence is hard to detect (but see \citealt{HughesEtal2011}), 

A further refinement which we will present in this paper is to quantify the role of the population of small dust grains. In resistive-MRI simulation, it is typical to assume that these particles are of (sub)micron-size and present in considerable amount (\eg\ GNT12 assumed a mass abundance of $10^{-3}$ in $0.1\ \mu$m-size grains). However, such small particles, being very sticky, should readily coagulate, as is well-known from theoretical and experimental studies \citep{ChokshiEtal1993,Blum2004}. Here, we will apply a coagulation model to solve for the effective abundance in small grains, thereby further reducing the available parameter space.

As our model chain is somewhat long, we first discuss, in \se{preview}, the relative motions among solid particles induced by turbulence for a wide size range. This is a typical outcome of our model, which we discuss in more detail in \se{model}. Results from our parameter study are presented in \se{results}. In \se{discuss} we address the question of the viability of runaway growth in the light of our findings. \Se{summary} presents our conclusions.

\section{Preview: Turbulent stirring across particle sizes}
\label{sec:preview}
\begin{figure*}[t]
  \centering
  \includegraphics[width=\textwidth]{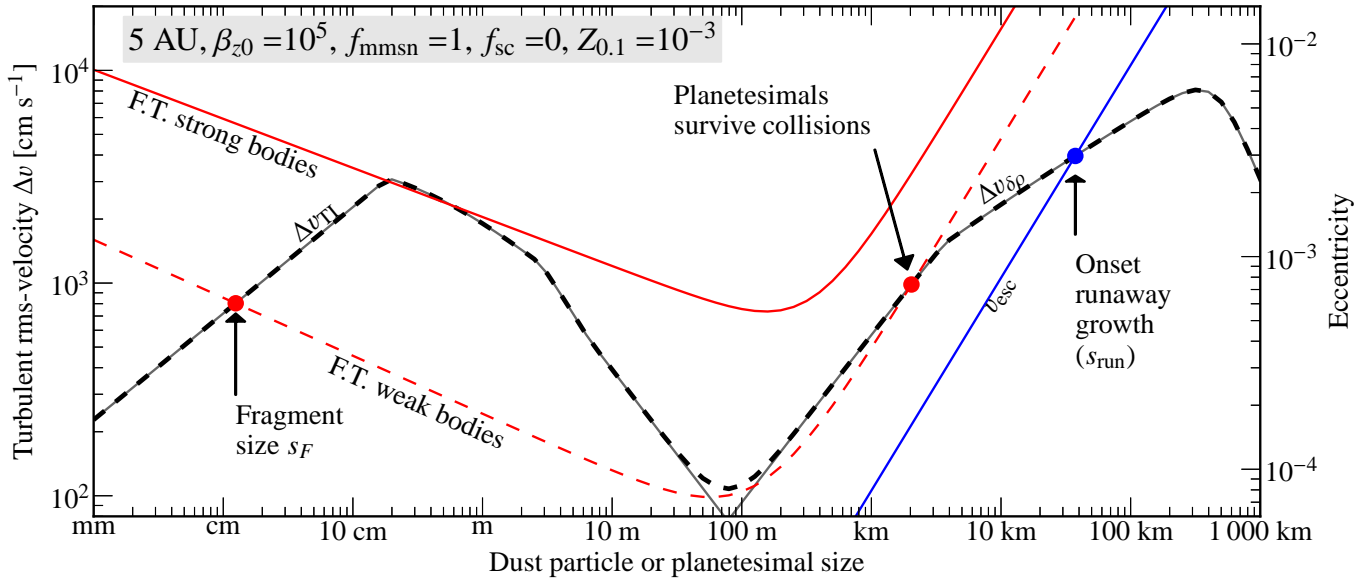}
  \caption{MRI turbulence-induced particle relative velocities $\Delta v$ in the midplane (dead zone) region of a protoplanetary disk at 5 AU for a \textit{fixed} mass abundance of $Z_\mathrm{0.1}=10^{-3}$ in $0.1\ \mu$m-size grains. The black dashed line shows the total turbulent-rms velocity between two particles of similar size $s$. The total $\Delta v$ consists of two components (gray curves): at large sizes it is governed by the turbulence-induced \textit{density fluctuations}, whereas at small sizes it is determined by the imperfect \textit{aerodynamical coupling} of particles to turbulent eddies. The point where $\Delta v$ starts to fall below the escape velocity $v_\mathrm{esc}$ (blue line) signifies the onset of runaway growth. Red solid and dashed lines give the fragmentation threshold (F.T.) above which similar-size particles will destroy themselves upon collisions.}
  %, adopting the prescription and parameters in \citet{StewartLeinhardt2009}. Therefore, if the material is weak (a good assumption when bodies experience multiple collisions), bodies in the cm--km size range fragment rather than accrete. }
  \label{fig:default}
\end{figure*}
Before presenting our model in detail, we show in \fg{default} its results in terms of the relative particle velocities among two particles of similar size (black-dashed curve). This is the result of our full model, outlined in \se{model}, for canonical disk parameters but \textit{without} considering dust coagulation. In models without dust coagulation the grains are assumed $0.1\ \mu$m in radius and present at an abundance of $10^{-3}$ with respect to the gas. This is denoted by $Z_{0.1} = 10^{-3}$ and a surface density of $\Sigma_{0.1} = Z_{0.1}\Sigma_\mathrm{gas}$. The standard disk parameters, described in detail in \se{disk}, are the following: a semi-major axis of $a=5$ AU, a plasma-beta parameter of $\beta_{z0}=10^5$ and the minimum-mass density for the surface density for the gas surface density $\Sigma_\mathrm{gas}$.  Our parameters closely match the standard model of GNT12 and, consequently, our results are very similar (but \fg{default} extends the size range considerably towards small sizes).\footnote{The field strength corresponding to our plasma beta paramter of $10^5$, $B_{z0}=4.2$ mG is somewhat lower than GNT12's D1.4b run (5.4 mG).  Also, GNT12 adopt a 10x higher ionizing flux contribution from short-lived radionuclides (see \se{disk}).} 

\Fg{default} illustrates that particle turbulent relative motions consist of two components, reflecting two different excitation mechanisms. For small particles it is eddy-driven turbulence; \ie\ small particles interact aerodynamically with the (fluctuating) turbulent velocity field. Due to their inertia they do not instantaneously couple to the motions of turbulent eddies, but lag their motion by a timescale $T_\mathrm{drag}$. This friction or stopping time is the time required for gas to damp the random velocity or eccentricity of particles.  This lag causes particles to acquire a relative motion with respect to the gas, and also with respect to themselves. However, for small dust particles, the particle-particle relative velocity is suppressed because their velocities are very coherent. This contrasts with the epicyclic motion of big bodies (planetesimals), for which it is usually fine to assume that their phase angles are random. But for small particles the situation is different; the motion of two (close) particles is that of the big eddy in which they are trapped.

At very small sizes, $\Delta v$ is small since solids are `glued' to the gas. (Indeed, for grains of size $s\lesssim\mu$m relative motions are driven by thermal [Brownian] motions, instead of turbulence, see \se{birnstiel}). Initially, the turbulence velocities increase linearly with size (this regime falls to the left of \fg{default}) and then switches to a square-root dependence on $s$. Turbulent inertia-driven velocities peak at dimensionless friction times $T_\mathrm{drag}\Omega = 1$ (at $s\simeq 10$ cm in \fg{default}) where particles obtain relative velocities of order $\delta v_\mathrm{mid}$. Thereafter, the turbulent inertia effect decreases as particles become too heavy to respond to the aerodynamical forcing of turbulent eddies.  

Rather, turbulent motions of bodies exceeding 100 m in radius are driven by gravitational interaction with density fluctuations. Here we obtain the equilibrium eccentricity by equating the stirring rate due to the density fluctuations in the gas by the damping rate due to gas drag (\se{T-scatter}).  Consequently, small planetesimals (say of 0.1--1 km size) obtain lower rms-eccentricities than bigger bodies, which is again a consequence of gas friction becoming less effective with increasing size. This increasing trend however stalls at a radius of $\approx$500 km and then declines rapidly due to tidal damping. 

The changes in the slopes seen at $2\times10^2$ cm and $3\times10^5$ cm reflect changes in gas drag law: in the first case the drag moves from the Epstein to the Stokes regime; in the second it becomes quadratic.
%\Fg{default} shows that Small particles are not immune to turbulence either. The mechanism by which small particles acquire their (relative) velocity differs, however, from that of large bodies. Whereas large bodies are forced gravitationally by density fluctuations, 
% From small to big sizes, therefore, the magnitude of turbulence-induced velocities changes (\fg{default}). 

In \fg{default} the solid and dashed red `strength curves' give an indication of the outcome of a collision between two bodies of similar size. Specifically, these curves mark the region where collisions are fragmentary (a larger velocity results in an object that is less massive) or accretionary (there is net mass gain). In calculating these velocity thresholds for fragmentation we have applied the velocity-dependent strength formulae of \citet{StewartLeinhardt2009} and shown two representative curves indicative of strong and weak bodies (see \se{turb-vel}). Although the material strength of bodies is uncertain and composition-dependent, we expect that for a collisionally-active system the `weak bodies' curve is more realistic.  Collisions between particles corresponding to sizes that lie in between the two red dots in \fg{default} are therefore fragmentary. The first point where the two curves intersect is denoted the fragmentation size $s_F$. Particle collisions with size $s<s_F$ are assumed to stick.

Finally, the blue line shows $v_\mathrm{esc}$; and the intersection (the blue dot) represents the size $s_\mathrm{run}$ (here $\approx$40 km) where planetesimals fulfill the condition for runaway growth. Note that this size is a factor of 10 larger than the fragmentation threshold (assuming weak materials). 

Although the model parameters corresponding to \fg{default} give rise to a dead zone, the turbulent forcing is nonetheless significant. A key goal of this work is to drop the assumption of a $Z_{0.1}=10^{-3}$ abundance in $0.1\ \mu$m-size particles and to replace it with a total dust abundance $Z_\mathrm{dust}$ in a \textit{distribution of grains} up to $s_F$, which arises due to dust coagulation (\se{birnstiel}). Other key parameters as $\beta_{z0}$ and the disk radius $a_0$ will likewise significantly affect the turbulent velocity curve and the intersections of this curve that determine the fragmentation size $s_F$ and the runaway growth size $s_\mathrm{run}$.
\begin{figure*}
  \centering
  \includegraphics[width=0.8\textwidth]{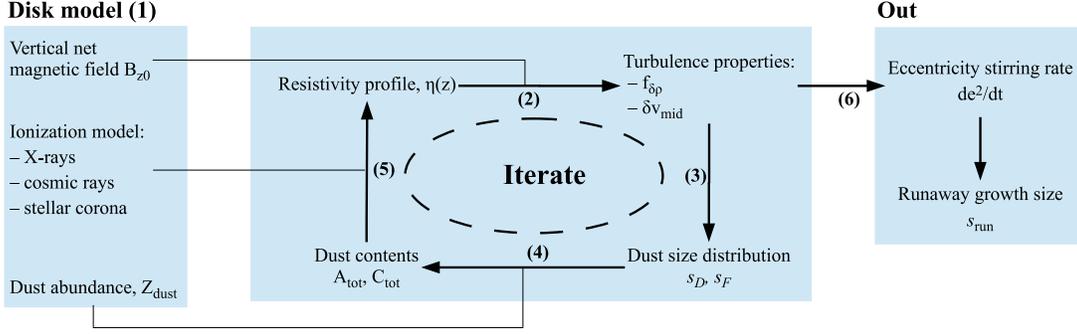}
  \caption{Flow chart of the model that provides the turbulence-induced stirring rate $de^2/dt$ and the runaway growth size $s_\mathrm{run}$. At its heart is an iteration procedure where the turbulence and the dust size distribution are consistently solved. In previous works we have detailed how the ionization fraction, which provides the resistivity profile $\eta(z)$, is calculated (step 5) and have provided predictor functions for the turbulent properties (step 2). From these, we calculate in step 3 the fragmentation size $s_F$ (step 3) and in present in step 4 a model for the dust size distribution. Step ($X$) is described in Section 3.$X$. }
  \label{fig:flowchart}
\end{figure*}

\section{Model}
\label{sec:model}
Our goal is to present a generic model that is able to quickly characterize the turbulent state at a local position in the disk and to describe how it excites solid bodies. Our strategy is to chain together several independently-developed semi-analytical recipes, in which the input of the one is the output of the other. These sub-models list (see \fg{flowchart}):
\begin{enumerate}
  \item A disk model. These are prescriptions for the surface density, disk radius, and strength of the vertical field which together determine the turbulent state of the disk. Other key parameters that affect the solution are the ionization model and the amount of the solid component that is in small particles (the dust component).
  \item A turbulence (dead zone) predictor model (OH11). Given a vertical resistivity profile for the gas, $\eta(z)$, the model obtains the statistical properties that characterize the turbulence, \ie\ the height of the dead zone, rms-gas velocity in the midplane, strength of the density fluctuations, \etc.
  \item A model for the turbulent velocities of small particles \citep{OrmelCuzzi2007} and for the strength of bodies \citep{StewartLeinhardt2009}. Together these provide an estimate of the upper size of the dust grain distribution, $s_F$, and the relative velocities of the dust particles in the turbulent-inertia regime.
  \item A coagulation-fragmentation model for the small dust size distribution \citep{BirnstielEtal2011}. We assume that small dust grains stick (coagulate), until they reach the fragmentation radius $s_F$, where they shatter and replenish the small grains. Under these conditions, a steady-state size distribution emerges. For simplicity, we assume that small dust particles do not coagulate on big bodies. (However, the implications of dust accretion can be obtained by varying the dust contents via the $Z_\mathrm{dust}$ parameter.) The dust model provides the total surface of dust per unit volume ($A_\mathrm{tot}$), which we also recast in terms of an equivalent abundance of grains ($Z_\mathrm{0.1,eqv}$).
  \item A charge-balance model for dust and gas \citep{Okuzumi2009}. From the dust size distribution, ionization properties, and assuming a dominant ionization species, we calculate the electron fraction of the gas as function of height. Consequently, we determines the resistivity of the gas $\eta(z)$.
  \item An improved stirring recipe for planetesimals (Paper II). From the turbulent properties computed in step (2) and the disk parameters, we have calculated the torques and the corresponding eccentricity excitation and diffusion rates on planetesimal bodies.
\end{enumerate}
Steps (2)--(5) should be iterated until convergence is achieved. Each sub-model ($X$) is detailed below in Section\ 3.$X$.

\subsection{Disk model}
\label{sec:disk}
\begin{deluxetable*}{llll}
  \tablecaption{\label{tab:model-pars}List of model parameters.}
  \tablehead{
  Parameter & Description & Values & Reference
  }
  \startdata
  $\Sigma_\mathrm{cr},\Sigma_\mathrm{xr}$ & Attenuation length cosmic rays, X-rays  & 96 and 8 g cm$^{-2}$ & \se{disk} \\
  $\beta_{z0}$        & Plasma beta midplane            & $10^4, \mathbf{10^5}, 10^6$ & \eq{plasma-beta} \\
  $\rho_{\bullet,D}$  & Internal density dust           & $3\ \mathrm{g\ cm}^{-3}$ \\
  $\rho_{\bullet,P}$  & Internal density planetesimals  & $2\ \mathrm{g\ cm}^{-3}$ \\
  $M_\star$           & Stellar mass                    & $1\ M_\odot$ \\
  $Z_\mathrm{dust}$   & Dust abundance                  & 0, $10^{-8},\dots \mathbf{10^{-3}}, 10^{-2}$ & \\
  $a_0$               & Disk radius                     & 1, \textbf{5}, 10 [AU]  & \Se{disk} \\
  $f_\Sigma$          & Disk mass                       & 0.1, \textbf{1}, 10     & \eq{sig-mmsn}\\
  $f_\mathrm{cr/xr/sr}$  & Control parameters for cosmic rays, X-rays, and radionuclide ionization rate & 1   & \Se{disk} \\
  $f_\mathrm{sc}$     & Control parameter for stellar corona protons & $\mathbf{0}, 1$  & \eq{ionization} \\
  $s_\mathrm{mon}$    & Dust grain minimum radius       & 0.1 \micr  & \se{birnstiel} \\
  $p_\mathrm{bm}$     & Slope size distribution Brownian motion regime  & $3/2$ & \Fg{Birnstiel} \\
  $p_\mathrm{turb}$   & Slope size distribution turbulent regime        & $1/4$ & \Fg{Birnstiel}
  \enddata
  \tablecomments{Multiple values indicate the parameter variation, with values in bold the default.}
%  The ionization rate is given in \eq{ionization}.}
\end{deluxetable*}

Throughout this paper we will assume that the gas surface density $\Sigma_\mathrm{gas}$ follows a power-law a function of disk radius $a$:
\begin{equation}
  \label{eq:sig-mmsn}
  \Sigma_\mathrm{gas} = 1.7\times10^3f_\Sigma\ \mathrm{cm^2\ g^{-1}} \left( \frac{a}{\mathrm{AU}} \right)^{-1.5},
\end{equation}
where $f_\Sigma$ is the enhancement of the surface density with respect to the minimum-mass solar nebula (MMSN) \citep{Weidenschilling1977i,HayashiEtal1985}. A gas-to-solid ratio of 100:1 by mass is assumed for simplicity, so that the surface density in solids, $\Sigma_\mathrm{solids}$, follows from \eq{sig-mmsn} simply by dividing by 100.  The temperature is given by
\begin{equation}
  T = 270\ \mathrm{K} \left( \frac{a}{\mathrm{AU}} \right)^{-0.5}
\end{equation}
and assumed isothermal in the the $z$-direction with scaleheight $H=c_s/\Omega$. The disk is assumed to be threaded by a magnetic field of magnitude $B_{z0}$, which will trigger the MRI. Instead of $B_{z0}$ we express the strength of the net field in terms of the plasma beta parameter $\beta_{z0}$, which is the ratio of the thermal to the magnetic energy at the midplane:
\begin{equation}
  \beta_{z0} = \frac{\rho_\mathrm{mid} c_s^2}{B_{z0}^2/8\pi}.
  \label{eq:plasma-beta}
\end{equation}

For the ionization rate $\zeta$ we follow GNT12 and \citet{TurnerDrake2009} and write:
\begin{eqnarray}
  \nonumber
  \zeta(z)
  &=&5\cdot10^{-18} \left(f_\mathrm{cr} +\frac{10^4f_\mathrm{sc}}{a_\mathrm{au}^2}\right) e^{-\frac{\Sigma_A}{\Sigma_\mathrm{cr}}} \left[1 +\left( \frac{\Sigma_A}{\Sigma_\mathrm{cr}} \right)^\frac{3}{4}\right]^{-\frac{4}{3}} +\ldots \\
  \nonumber
  &&+ 2.6\cdot10^{-15} f_\mathrm{xr} \frac{ e^{-\Sigma_A/\Sigma_\mathrm{xr}} } {a_\mathrm{au}^2} +\ldots \\
  &&+ 3.7\cdot10^{-19} f_\mathrm{sr}
  \label{eq:ionization}
\end{eqnarray}
%and $\Sigma_B=\Sigma_\mathrm{gas}-\Sigma_A$ are
where $\zeta$ is in units of $\mathrm{s}^{-1}$, $\Sigma_A$ the surface density above a height $z$, $\Sigma_\mathrm{cr} = 96\ \mathrm{g\ \mathrm{cm^{-2}}}$ and $\Sigma_\mathrm{xr}=8.0\ \mathrm{g\ cm^{-2}}$ attenuation lengths for cosmic rays and X-rays, and $f_\mathrm{cr}$, $f_\mathrm{xr}$, $f_\mathrm{sr}$ control parameters for the contributions due to cosmic rays, X-rays, and short-lived radionuclides. 
In \eq{ionization} `$\ldots$' implies that the contribution from the lower disk ($\Sigma_B = \Sigma_\mathrm{gas} -\Sigma_A$) must be added.  GNT12 used $f_\mathrm{cr}=f_\mathrm{xr}=1$ and $f_\mathrm{sr}=10$. Here, we take $f_\mathrm{sr}=1$. On the other hand, following \citet{TurnerDrake2009}, we do account for the possibility of a large contribution from protons originating from the stellar corona when $f_\mathrm{sc}\neq0$.  However, this contribution is rather uncertain (for example, the stellar protons may be channeled back to the star; see the arguments outlined in \citealt{TurnerDrake2009}); GNT12 did not account for these stellar protons. Our default here is to omit this contribution (\ie\ $f_\mathrm{sc}=0$), but we will also run models that include this term ($f_\mathrm{sc}=1$). In this way we test the sensitivity of the results against a sharp increase in the ionizing flux.

%In that case we follow the prescription of \citet{TurnerDrake2009} and use the same attenuation formula as for cosmic rays. \citet{TurnerDrake2009} employ assumptions that cause a large contribution from the stellar corona (because they wish to obtain small dead zones) -- assumptions that are rather uncertain. The more conservative approach is to neglect this contributions ($f_\mathrm{sc}=0$ as GNT12 did), which is our default. Here, we simply want to test the sensitivity of our results agains a large ionization source. 

Following Paper I it is assumed that the dominant ionization species is $\mathrm{H}_3^+$ with an gas-phase recombination rate coefficient of $6.7\times10^{-18} (T/\mathrm{300\ K})^{-0.5} \mathrm{cm^3\ s}^{-1}$ \citep{McCallEtal2004}.

\subsection{Turbulence predictor model}
\label{sec:predict}
OH11 provide simple scaling relationships for the turbulent properties that characterize an Ohmic-resistive MRI-active disk. The heart of the model is to compute a set of scaleheights, $H_\mathrm{idl0}\ge H_\mathrm{\Lambda0}\ge H_\mathrm{res0}$, which follow from the disk parameters defined above and the resistivity profile (see Equations (11)--(13) of OH11).  Crudely, these scaleheights correspond, respectively, to the scale where the MRI turbulence becomes ideal, resistive, and dead. Using these scaleheights and aided by their simulations OH11 subsequently formulated predictor functions (recipes) for the emergent quantities of the turbulence. For example, $\alpha_\mathrm{core}(H_{\Lambda0}, H_\mathrm{res0}, B_{z0})$ gives the level of turbulent activity (stresses) in the midplane regions (Equation (28) of OH11) and $\delta v_\mathrm{mid}$ the rms-turbulent gas velocity (Equation (47) of Paper I):  
\begin{equation}
  \delta v_\mathrm{mid}
  = \sqrt{1.1 {\cal L}\alpha_\mathrm{core}} c_s
  \label{eq:v-mid}
\end{equation}
where ${\cal L}$ is a flux limiter -- a correction term that becomes less than unity for strong fields.  We refer to OH11 and Paper I for further details. In addition, we have, in Paper I, augmented the model with a prescription for the behavior of solid bodies, as they interact gravitationally with the gas density fluctuations that the turbulence produces. This is discussed in \se{T-scatter}.

%but let us briefly comment on the physics behind the expressions appearing in this equation. 

\subsection{Turbulent relative velocity for small particles and characteristic sizes $s_D, s_F$}
\label{sec:turb-vel}
The interaction of small particles with turbulent eddies is determined by the aerodynamic properties of the particles, quantified by their friction times ($T_\mathrm{drag}$). In turbulence, the ratio of the friction time to the driving scales of the turbulence -- at both the high end and the low end of the spectrum -- matter. Here, we take the inverse orbital frequency $\Omega^{-1}$, as the turnover time of the largest eddies and define the Stokes number as $\mathrm{St} = T_\mathrm{drag}\Omega$. \citet{VoelkEtal1980} introduced a framework to calculate particle relative velocities assuming hydrodynamic turbulence characterized by a Kolmogorov cascade.  This model has been refined by subsequent works \citep{MarkiewiczEtal1991,CuzziHogan2003,PanPadoan2010}. We refer to these eddy-driven relative velocities as $\Delta v_\mathrm{TI}$ (TI = Turbulent Inertia) and we adopt the closed-form expressions of \citet{OrmelCuzzi2007}.

The relative velocity between similar-size particles in the intermediate size regime -- valid when the particle friction time falls between $\Omega^{-1}$ and the turnover time of the smallest turbulent eddies -- is $\Delta v_\mathrm{TI} \approx 1.4 \mathrm{St}^{1/2} \delta v_\mathrm{mid}$ \citep{OrmelCuzzi2007}. It increases with increasing particle size (or Stokes number $\mathrm{St}$) as particles couple more loosely to the gas. At some point, then, collisional energies will be large enough for particles to fragment.  The corresponding size is referred to as the \textit{fragmentation threshold} $s_F$: above it collisions between (similar-size) particles result in fragmentation; below it, they stick. Specifically, we obtain $s_F$ by equating the specific collisional energy, which is $(\Delta v_\mathrm{TI})^2/8$ for particles of equal size, to a material and velocity dependent threshold $Q_\mathrm{RD}^\ast$; \ie\
\begin{equation}
  \frac{[\Delta v_\mathrm{TI}(s_F)]^2}{8} = Q_\mathrm{RD}^\ast(s_F, \Delta v_\mathrm{TI})
  \label{eq:sF-def}
\end{equation}
\citep{StewartLeinhardt2009}. For the strength curve $Q_\mathrm{RD}^\ast$ we copy the parameters of \citet{StewartLeinhardt2009} corresponding to weak bodies (the dashed line in \fg{default}).  (Fragmented) dust below $s=s_F$ will start to re-coagulate. The result is that the size distribution at every $s\le s_F$ is balanced by losses (due to coagulation to larger radii) and gains (coagulation from smaller particles and fragmenting collisions involving $s_F$-particles). As the evolution timescales of small particles are short, a (quasi) steady-state is reached.

Another critical radius, which we refer to as the dust size $s_D$, is the radius where the relative velocity induced by turbulence equals those due to thermal (Brownian) motions, \ie\ that $\Delta v_\mathrm{TI}(s_D) = \Delta v_\mathrm{BM}(s_D)$, where $\Delta v_\mathrm{BM} \approx \sqrt{k_B T/m_D}$ with $k_B$ Boltzmann's constant and $m_D$ the mass corresponding to $s_D$. According to Eq.\ (37) of \citet{BirnstielEtal2011}:
\begin{eqnarray}
  \label{eq:s-g}
  s_D 
  &=& \left[ \frac{8\Sigma_\mathrm{gas}}{\pi\rho_{\bullet}} \left( \frac{1}{\alpha_T^2 \mathrm{Re}} \right)^{1/4} \sqrt{\frac{m_\mu}{4\pi^2\rho_{\bullet}}} \right]^{2/5} \\
  \nonumber
  &\approx& 1.2\ \mu\mathrm{m}\ \left( \frac{\Sigma_\mathrm{gas}}{10^3\ \mathrm{g\ cm}^{-2}} \right)^{2/5} \left( \alpha_T^2 \mathrm{Re} \right)^{-1/10} \left( \frac{\rho_\bullet}{3\ \mathrm{g\ cm^{-3}}} \right)^{-3/5},
\end{eqnarray}
where $m_\mu$ is the mean molecular mass of the gas (assumed to be 2.3amu), $\alpha_T=(\delta v_\mathrm{mid}/c_s)^2$ a dimensionless measure of the gas rms-velocity at the midplane regions, and $\mathrm{Re}$ the Reynolds number, defined as $\mathrm{Re}=\alpha_T c_s H/\nu_\mathrm{mol}$ with $\nu_\mathrm{mol}$ the molecular viscosity. Thus, given the state of the turbulence as provided by the predictor for $\delta v_\mathrm{mid}$, the radius $s_D$ follows from \eq{s-g}.

One caveat pertains the validity of the \citet{OrmelCuzzi2007} expressions in dead zones, where the nature of the fluctuating gas motions is due to dissipating sound waves rather than vortical turbulence (see \eg\ \citealt{HeinemannPapaloizou2012} or GNT12). While the \citet{OrmelCuzzi2007} expressions assume a Kolmogorov-like cascade, the power of the sound waves might be more concentrated at a large scale (\ie\ at frequency $\sim$$\Omega$). In that case, the OC07 expressions would overestimate the relative velocity for particles obeying $\mathrm{St}<1$, resulting in a lower fragmentation threshold size $s_F$, which in turn implies that the size of the dead zone has been overestimated. Such non-Kolmogorov turbulence will therefore reinforce our conclusion that coagulation efficiently depletes small dust grains.

%For the OC07 expressions to hold in dead zones, a Kolmogorov-like spectrum must exist -- at least in the temporal domain. This, we believe is reasonable (\ie\ the motions of particles smaller than $\mathrm{St}<1$ will be correlated in a similar fashion), although more effort is desired.}

\subsection{A model for the dust size distribution}
\label{sec:birnstiel}
%We consider a quasi steady-state dust model where the dust size distribution is determined from the balance between coagulation and fragmentation. Specifically, 
The dust properties of the distribution affect the ionization balance of the gas.  In particular, in order to solve for the charge balance, we must obtain the total dust surface area per unit volume ($A_\mathrm{tot}$) and the total size per unit volume ($C_\mathrm{tot}$). Thus, we need a model for the dust size distribution and calculate its moments. 

Apart from $s_F$ and $s_D$, the size distribution is further characterized by a cut-off size at the smallest (monomer grain) radius $s_\mathrm{mon}$.  We follow the model of \citet{BirnstielEtal2011} to find the power-law exponent of the size distribution, see \fg{Birnstiel}. In this model, dust particles coagulate until the fragmentation radius $s=s_F$, beyond which particles fragment according to a certain size distribution. For the sake of simplicity, we only retain its main features, see \fg{Birnstiel}. 
\footnote{The ignored effects, described in \citet{BirnstielEtal2011} are: variations in the scaling of $\Delta v_\mathrm{TI}$ with friction time; the vertical stratification of particles; and a pileup of surface density at a radius $s_F$ due to boundary effects.}
%Apart from a fragmentation size $s_F$, %\footnote{$\alpha_T$, a proxy for the gas motions at the midplane, is not the same as $\alpha_\mathrm{core}$, which is a proxy for the turbulent stresses. See Paper I for details.}

\begin{figure}[t]
  \includegraphics[width=85mm]{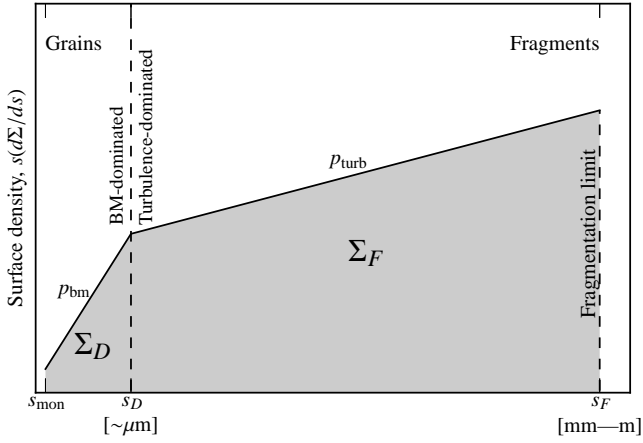}
  \caption{Sketch of the dust component of the model. A coagulation/fragmentation equilibrium is assumed. Dust particles coagulate until $s=s_F$, the fragmentation threshold, where turbulence relative motions are large enough for particles to fragment. The size distribution is characterized by two power-law indices depending on the mechanism that drives relative motions \citep{BirnstielEtal2011}. The mass of the distribution is dominated by particles near the largest fragments size $s_F$, but the surface area ($A_\mathrm{tot}$) is dominated by particles around the `dust size' $s_D$, below which small particles are efficiently removed by Brownian motion.}
  \label{fig:Birnstiel}
\end{figure}
The size distribution is modeled as a power-law, characterized by an exponent $p$, defined such that
\begin{equation}
  \label{eq:p-def}
  \frac{d\Sigma}{d\log s} \propto s^{p}
\end{equation}
measures the amount of mass ($\Sigma$) in a logarithmic size bin.  Based on the size-dependence of the velocity and the spectrum of particles fragmented at $s_F$, \citet{BirnstielEtal2011} provide expressions for $p$ consistent with steady-state.  As Brownian motion and turbulence exhibit quite different dependences on particle size (for Brownian motion the velocity scales as $s^{-3/2}$ while for turbulence it scales linearly with size), $p$ naturally changes at the point $s_D$.  The dust distribution is therefore characterized by two exponents: $p_\mathrm{bm}$ in the BM-regime, and $p_\mathrm{turb}$ in the turbulent regime (see \fg{Birnstiel}). Following \citet{BirnstielEtal2011} (see their Table\ 3) we take $p_\mathrm{bm}=3/2$ and $p_\mathrm{turb}=1/4$.
\footnote{Like \citet{BirnstielEtal2011} we have assumed that the fragments are re-distributed by a power-law of $p_\mathrm{frag}=1/2$. Note that the caption of Table\ 3 of \citet{BirnstielEtal2011} suggests a different power-law index than \eq{p-def}. But this is erroneous; their exponent is defined as in \eq{p-def}.}

The following discussion assumes that $s_\mathrm{mon} \ll s_D \ll s_F$. If the total surface density in particles of radius $s_\mathrm{mon}\le s \le s_D$ is $\Sigma_D$ and $\Sigma_F$ for particles of radius $s_D \le s \le s_F$ then the size distribution function (\eqp{p-def}) becomes:
\begin{equation}
  \frac{d\Sigma}{d\log s}
  = \left\{
  \begin{array}{ll}
    \displaystyle
    \frac{3\Sigma_D}{2} \left( \frac{s}{s_D} \right)^{3/2} & (s_\mathrm{mon} \le s \le s_D) \\[5mm]
    \displaystyle
    \frac{ \Sigma_F}{4} \left( \frac{s}{s_F} \right)^{1/4} & (s_D \le s \le s_F)
  \end{array}
  \right.
  \label{eq:dSigds}
\end{equation}
For the adopted values of the power-law exponents, \eq{dSigds} implies that the surface density is dominated by particles of size $s_F$, the total surface area of the dust ($A_\mathrm{tot}$) by particles around the dust radius $s_D$ and the total size per unit volume ($C_\mathrm{tot}$) by particles around the monomer radius $s_\mathrm{mon}$.
%, it follows that all critical radii ($s_\mathrm{mon}$, $s_D$ and $s_F$) affect the resistivity of the gas.  

Expressions for $A_\mathrm{tot}$ and $C_\mathrm{tot}$ also depend on the height $z$ above the midplane. Let us therefore consider volume densities $\rho_\mathrm{dust}(z)$ instead of integrated surface densities ($\Sigma$). When it is assumed that the dust particles follow the same vertical distribution as the gas, we can simply replace $\Sigma$ by $\rho_\mathrm{dust}(z)$, $\Sigma_D$ by $\rho_D(z)$, and $\Sigma_F$ by $\rho_F(z)$. For the \textit{number density} distribution $dn/ds$ we further divide by the mass of a particle, $m=4\pi \rho_\bullet s^3/3$. Thus,
\begin{equation}
  \frac{dn}{d\log s} 
  = \left\{
  \begin{array}{ll}
    \displaystyle
    \frac{9\rho_D}{8\pi \rho_\bullet s_D^3} \left( \frac{s}{s_D} \right)^{-3/2} & (s_\mathrm{mon} \le s \le s_D) \\[5mm]
    \displaystyle
    \frac{3\rho_F}{16\pi\rho_\bullet s_F^3} \left( \frac{s}{s_F} \right)^{-11/4} & (s_D \le s \le s_F)
  \end{array}
  \right.
\end{equation}
gives the particle size distribution per unit volume.  The total surface area then becomes
\begin{eqnarray}
  \nonumber
  A_\mathrm{tot} 
  &\equiv& \int \frac{dn}{ds} \pi s^2\ ds
  \approx \frac{18\rho_D}{8 \rho_\bullet s_D} +\frac{4\rho_F}{16\rho_\bullet s_F} \left( \frac{s_F}{s_D} \right)^{3/4} \\
  &=& \frac{5\rho_F}{8\rho_\bullet s_D} \left( \frac{s_F}{s_D} \right)^{-1/4}
  \label{eq:Atot}
\end{eqnarray}
where in the last step we used that the size distribution is continuous at $s=s_D$:
\begin{equation}
  \frac{\Sigma_D}{\Sigma_F}
  = \frac{\rho_D}{\rho_F}
  = \frac{1}{6} \left( \frac{s_F}{s_D} \right)^{-1/4},
\end{equation}
which again assumes $s_\mathrm{mon} \ll s_D \ll s_F$.

It is instructive to compare the value of $A_\mathrm{tot}$ for the distribution (\eqp{Atot}) with that in case of a monodisperse grain population of radius $s_\mathrm{mon}$ and surface density $\Sigma_0$. In that case
\begin{equation}
  A_\mathrm{tot-mono}
  = \frac{3\rho_\mathrm{mon}(z)}{4s_\mathrm{mon}\rho_\bullet}.
\end{equation}
Equating this expression to \eq{Atot}, we define the \textit{surface area-equivalent 0.1 $\mu$m dust surface density}, \ie\ the surface density in $0.1$ \micr\ grains which amounts to the same $A_\mathrm{tot}$ as that of the steady state dust distribution:
\begin{equation}
  \label{eq:sig0eff}
  \Sigma_\mathrm{0.1,eqv} 
  \equiv \frac{5}{6} \Sigma_F \left( \frac{s_F}{s_D} \right)^{-1/4} \left( \frac{s_D}{0.1\ \micr} \right)^{-1},
\end{equation}
where we took $s_\mathrm{mon}=0.1\ \mu$m.  Similarly, we define an equivalent abundance $Z_\mathrm{0.1,eqv} = \Sigma_\mathrm{0.1,eqv}/\Sigma_\mathrm{gas}$ as the abundance by mass in 0.1 \micr\ size grains which amounts to the same $A_\mathrm{tot}$ as the distribution.  Note that the prefactor of 5/6 in \eq{sig0eff} is an artifact of the assumption that $s_\mathrm{mon} \ll s_D \ll s_F$. Since both terms in the brackets of \eq{sig0eff} are $\le 1$, the surface area-equivalent surface density in very small grains is always smaller than the total surface density in dust, which is dominated by $\Sigma_F$. A larger $s_D$ acts to decrease $\Sigma_\mathrm{0.1,eqv}$, because coagulation by Brownian motion becomes more important. A larger $s_F$ locks more mass in bigger particles. Both effects imply that the stronger the turbulence, the larger $\Sigma_\mathrm{0.1,eqv}$ becomes.

\Eq{sig0eff} is useful to interpret our results in terms of a single grain radius $s_\mathrm{mon}$ (here fixed at $0.1$ \micr) as the ratio $\Sigma_\mathrm{0.1,eqv}/\Sigma_F \approx Z_\mathrm{0.1,eqv}/Z_\mathrm{dust}$ is a measure for the reduction of the dust (surface area) due to coagulation.  These considerations suggest that MRI simulations modeling resistivity effects can keep using a single grain size, but in order to mimic the effects of dust distribution, must reduce its abundance accordingly as otherwise the dust surface area will be unrealistically high (alternatively, one could choose a large grain size). 

\subsubsection{Note on assumptions regarding the dust distribution model}
In closing this section, we comment on some of the assumptions made in obtaining $Z_\mathrm{0.1,eqv}$. For example, we assumed the intermediate-mass regime for the turbulent velocity, which requires $T_\mathrm{drag}(s_F)\Omega<1$. This, it turns out, is always satisfied. Furthermore, we assumed that the particles have the same scaleheight as the gas. For this assumption to hold one requires $T_\mathrm{drag}\Omega \lesssim \alpha_T$ \citep{CuzziEtal2005}, which is not always satisfied. However, we think that the implications are limited. Firstly, although the size distribution is more accurately described by a three-piece (or even a four-piece when we also account for changes in the turbulent velocity; see \citealt{BirnstielEtal2011}) function of size, it will not alter the fact that most of the mass is in particles around $s_F$. Secondly, small grains around the dust size $s_D$, which dominates $A_\mathrm{tot}$, are always distributed with the same scaleheight as the gas. Thus, the resistivity profile, $\eta(z)$, should not be much affected.

Another assumption was that dust fragmentation occurs only among $s_F$ particles. Conceivably, smaller particles could diffuse to the MRI-active regions (\eg\ \citealt{CarballidoEtal2011}) where they are much more likely to experience fragmenting collisions due to the higher turbulent gas velocity and lower gas density. However, we also believe these effects are limited as (i) the collision rate, being proportional to the square of the density, drops substantially for $z>H$; and (ii) due to the increase in $Q_\mathrm{RD}^\ast$ with decreasing size (see \fg{default}), no large variations in $s_F$ are expected. Concerning the model, larger sources of uncertainty pertain the fragmentation law (\ie\ the $Q_d^\ast(s)$ as function of size), the power-law index $p_\mathrm{frag}$ of the collision products in fragmenting collisions, as well as our neglect of porous aggregation. These factors can be addressed in principle, but are beyond the scope of the present study.

%Conversely, \eq{sig0eff} also implies that not every choice of $Z_\mathrm{0.1,eqv}$ corresponds to a realistic dust size distribution.
%We find that the a single grain size with $\Sigma_\mathrm{0.1}=\Sigma_\mathrm{0.1,eqv}$ results in the same gas ionization fraction than the 
%charge  that are indistinguishable from the results based on the raw $A_\mathrm{tot}, C_\mathrm{tot}$ (\eqs{Atot}{Ctot}). \com{Maybe therefore slim the discussion of Ctot?}
%\footnote{For $A_\mathrm{tot}$ this result is not surprising since $\Sigma_\mathrm{0.1,eqv}$ is defined in terms of $A_\mathrm{tot}$. Any error should therefore be attributed to $C_\mathrm{tot}$ (\eq{Ctot}), which does not follow from $\Sigma_\mathrm{0.1,eqv}$ and $s_\mathrm{mon}$.}

\subsection{A charge balance model for the dust and the gas}
The next step is to calculate the ionization fraction of the gas, $x_e$, for which we use the dust-grain charge model of \citet{Okuzumi2009}. The dust mainly affects the outcome by the total surface area $A_\mathrm{tot}$ and (weakly) by the total capacitance (the total size per unit volume), $C_\mathrm{tot}$. The latter dependence arises because the average grain charge of a dust grain is proportional to the grain radius \citep{Okuzumi2009}. Assuming \eq{dSigds}, we can readily solve for $C_\mathrm{tot}$:
\begin{equation}
  C_\mathrm{tot}
  = \int \frac{dn}{ds} s\ ds 
% \approx \frac{18\rho_D}{8\pi\rho_\bullet s_D^2} \left( \frac{s_D}{s_\mathrm{mon}} \right)^{1/2} \\
  = \frac{3\rho_F}{8\pi\rho_\bullet s_D^2} \left( \frac{s_D}{s_\mathrm{mon}} \right)^{1/2} \left( \frac{s_F}{s_D} \right)^{-1/4}.
  \label{eq:Ctot}
\end{equation}
Note that $C_\mathrm{tot}$ only modestly depends on $s_\mathrm{mon}$, while $A_\mathrm{tot}$ does not. Thus, the choice for $s_\mathrm{mon}$ (which is a parameter) will not much affect the conclusions of this work.

\citet{Okuzumi2009} solves the ionization balance in terms of a dimensionless parameter, $\Theta$, which depends on grain properties ($A_\mathrm{tot}$, $C_\mathrm{tot}$), gas properties, and the ionization rate, $\zeta$ (Equation (31) of \citealt{OkuzumiEtal2009}). The value of $\Theta$ reflects the dominant carriers of negative charge: free electrons (for which $\Theta\gg 1$) or negatively-charged dust (for which $\Theta\ll 1$). In our case, it turns out that dust coagulation drives the solution towards $\Theta\gg1$, the ion-electron plasma limit. In this limit, the ionization fraction of the gas $x_e$ becomes insensitive to $C_\mathrm{tot}$ \citep{Okuzumi2009}; and our description in terms of an effective surface density (\eqp{sig0eff}) becomes exact.

We solve for the ionization fraction as function of height, $x_e = x_e(z)$ (Equation (28) of \citealt{Okuzumi2009}). The ionization fraction of the gas in turn determines the resistivity profile $\eta(z)$ 
\begin{equation}
  \eta(z)
  =
  \frac{234 \sqrt{T}}{x_e(z)}
\end{equation}
\citep{BlaesBalbus1994}. This completes the iteration cycle. With the updated $\eta(z)$, we can now go back to \se{predict} and iterate steps (2)--(5) until convergence is achieved.

A key parameter in \citet{Okuzumi2009}'s charge-balance model is the choice for the (dominant) ionization species, as it determines (among other) the recombination rate.  Following paper I, we reserve this role for the H$_3^+$ ion. However, in the midplane charge-exchange reactions will cause Mg$^+$ to become the dominant ionization species \citep{IlgnerNelson2006,AdamkovicsEtal2011}. To estimate the uncertainty arising from the choice of a single ionization species, we have rerun our results assuming that Mg$^+$ is the dominant ion. We generally found the same trends emerging, although $s_\mathrm{run}$, for example, is somewhat larger and more sensitive to the dust abundance $Z_\mathrm{dust}$.

%As a result, the solutions
%Correspondingly, the ionization fraction of the gas reads
%\begin{equation}
%  x_e
%  =
%  \frac{\zeta}{s_e u_e A_\mathrm{tot}} \exp \Gamma_\infty,
%%\end{equation}
%where $s_e$ is a sticking coefficient for the electrons, $u_e$ the thermal velocity of the gas, and $\Gamma_\infty$ an order-of-unity factor, which depends on the properties of the ions and electrons.
%However, it turns out that at $z = H_\mathrm{\Lambda0}, H_\mathrm{res0}$ the charge balance equation follows the ion-electron plasma limit. 
%However, the ionization fraction of the gas is determined by the total surface area per unit volume $A_\mathrm{tot}$ and the total size per unit volume $C_\mathrm{tot}$ \citep{Okuzumi2009,OkuzumiEtal2011}, which .   

\subsection{Turbulence-induced scattering model for planetesimals}
\label{sec:T-scatter}
In Paper I we have presented the model for the stochastic behavior of solid bodies in phase space (semi-major axis and eccentricity) as function of disk parameters. For the eccentricity stirring we obtained (Equation 49 of Paper I):
\begin{equation}
  \label{eq:e-tidf}
  \left( \frac{de^2}{dt} \right)
  = 2D_e
  = \frac{0.94{\cal L}\alpha_\mathrm{core}}{(1+4.5H_\mathrm{res0}/H)^2} \left( \frac{\Sigma_\mathrm{gas} a_0^2}{M_\star} \right)^2 \Omega \\
  \equiv f_\mathrm{\delta\rho}^2 q_\mathrm{gas}^2 \Omega,
  \label{eq:De-def}
\end{equation}
%For simplicity, we will in this work write the first term on the RHS as $f_{\delta\rho}^2$ and write the stirring rate as
%\begin{equation}
%  \left( \frac{de^2}{dt} \right)_{\delta \rho} 
%\end{equation}
where in the last step we defined $f_\mathrm{\delta\rho} = \sqrt{0.94{\cal L}\alpha_\mathrm{core}} /(1+4.5H_\mathrm{res0}/H)$ and $q_\mathrm{gas} = \Sigma_\mathrm{gas} a^2/M_\star$. In the case of ideal-MRI $f_{\delta\rho} \approx \alpha_\mathrm{core}^{1/2} \approx \delta\rho/\rho\approx 0.1$, but \eq{De-def} contains two key correction factors that reduce $f_{\delta\rho}$. Firstly, it was found that the density fluctuations are suppressed at high values of the magnetic field. In \eq{De-def} this effect is accounted for by the flux-limiter ${\cal L}$ (Equation (41) of Paper I), which becomes less than unity when fields become strong.  Secondly, we included a correction for the geometric distortion of the density waves in the case of dead zones ($H_\mathrm{res0}\neq0$). As first reported by GNT12 the density waves get sheared out as they travel from the MRI-active layers to the midplane.  The term in the denominator of \eq{De-def}, $1+4.5H_\mathrm{res0}/H$, accounts for this effect. In dead zones, therefore, $f_{\delta\rho}\ll0.1$.

We assume that the eccentricity is damped by gas drag and (for large planetesimals) tidal damping:
\begin{equation}
  \label{eq:e-damp}
  \left( \frac{de^2}{dt} \right)_\mathrm{damp} 
  \simeq \frac{2e^2}{T_\mathrm{damp}},
\end{equation}
where $T_\mathrm{damp}^{-1} = T_\mathrm{drag}^{-1} +T_\mathrm{tidal}^{-1}$ with $T_\mathrm{tidal}$ the timescale for tidal damping \citep{TanakaWard2004} and $T_\mathrm{drag}$ the gas drag timescale of particles, 
\begin{equation}
  T_\mathrm{drag} 
  = \frac{8s_p\rho_\bullet}{3C_D\rho_\mathrm{gas} v_\mathrm{gas}}
  \label{eq:Tdrag}
\end{equation}
with $C_D$ the drag constant, and $v_\mathrm{gas}$ the gas-particle relative velocity. A relative velocity arises due to the eccentric motions of a body as well as the radial pressure gradient of the gas, which causes it to rotate lower than Keplerian by a magnitude $v_\mathrm{hw}$ (the headwind). Combining these effects we approximate $v_\mathrm{gas} \approx ea\Omega +v_\mathrm{hw}$ and take $v_\mathrm{hw}=30\ \mathrm{m\ s}^{-1}$.  Since the drag constant $C_D$ depends on $v_\mathrm{gas}$ and $v_\mathrm{gas}$ on eccentricity, an iterative approach is generally needed to solve for the equilibrium eccentricity, \ie\ the value of $e$ that satisfies $(de^2/dt)_\mathrm{damp} = (de^2/dt)_\mathrm{\delta\rho}$. 

In  this paper, we will focus on the point where the runaway growth condition, $\Delta v < v_\mathrm{esc}$, is satisfied. The corresponding radius $s_\mathrm{run}$ is marked by the blue dot in \fg{default}.  We find $s_\mathrm{run}$ from the runaway growth condition, $\Delta v_{\delta\rho} = v_\mathrm{esc}$ where $\Delta v_{\delta\rho}$ is obtained by equating \eq{e-tidf} to \eq{e-damp} (as we did in \fg{default}).  

% AU & & & & $H_g$ & & cm & $\mu$m & km \\
\begin{deluxetable*}{lllllllllll}
  \tablewidth{\textwidth}
  \tablecaption{\label{tab:output}Several output quantities corresponding to runs where $Z_\mathrm{dust}=10^{-3}$.}
  \tablehead{
  $a$ [AU] & $f_\Sigma$ & $\beta_{z0}$ &$f_\mathrm{sc}$ &$\alpha_T$ &$f_{\delta\rho}$ &$H_\mathrm{res0}/H$  & $s_D [\micr]$ & $s_F [\mathrm{cm}]$ & $Z_\mathrm{0.1,eqv}$ & $s_\mathrm{run}$ [km]
  }
  \startdata
{$1$} & {$0.1$} & {$10^{4}$} & {$0$} & {$2.5(-2)$} & {$1.8(-2)$} & {$1.5$} & {$0.15$} & {$1.1(-1)$} & {$3.4(-5)$} & {$1.8(1)$} \\
{$1$} & {$0.1$} & {$10^{5}$} & {$0$} & {$2.1(-3)$} & {$4.5(-3)$} & {$1.9$} & {$0.31$} & {$5.6(-1)$} & {$1.7(-5)$} & {$1.7$} \\
{$1$} & {$0.1$} & {$10^{6}$} & {$0$} & {$1.6(-4)$} & {$1.1(-3)$} & {$2.3$} & {$0.66$} & {$2.9$} & {$7.2(-6)$} & {$2.9(-3)$} \\
{$1$} & {$1$} & {$10^{4}$} & {$0$} & {$9.4(-3)$} & {$7.9(-3)$} & {$2.3$} & {$0.39$} & {$9.3(-1)$} & {$1.3(-5)$} & {$2.7(1)$} \\
{$1$} & {$1$} & {$10^{5}$} & {$0$} & {$1.4(-3)$} & {$2.9(-3)$} & {$2.5$} & {$0.68$} & {$3.1$} & {$6.9(-6)$} & {$5.9$} \\
{$1$} & {$1$} & {$10^{6}$} & {$0$} & {$1.2(-4)$} & {$7.6(-4)$} & {$2.8$} & {$1.4$} & {$8.4$} & {$3.4(-6)$} & {$5.0(-1)$} \\
{$1$} & {$10$} & {$10^{4}$} & {$0$} & {$8.7(-4)$} & {$1.9(-3)$} & {$2.9$} & {$1.6$} & {$3.9$} & {$3.9(-6)$} & {$1.9(1)$} \\
{$1$} & {$10$} & {$10^{5}$} & {$0$} & {$9.2(-4)$} & {$1.9(-3)$} & {$3.1$} & {$1.6$} & {$3.8$} & {$3.9(-6)$} & {$1.9(1)$} \\
{$1$} & {$10$} & {$10^{6}$} & {$0$} & {$1.0(-4)$} & {$5.8(-4)$} & {$3.3$} & {$3.1$} & {$9.3$} & {$2.0(-6)$} & {$2.6$} \\
{$5$} & {$0.1$} & {$10^{4}$} & {$0$} & {$5.7(-2)$} & {$2.2(-1)$} & {$0.0$} & {$0.10$} & {$2.3(-2)$} & {$5.6(-5)$} & {$6.8(2)$} \\
{$5$} & {$0.1$} & {$10^{5}$} & {$0$} & {$5.6(-3)$} & {$7.0(-2)$} & {$0.0$} & {$0.11$} & {$1.0(-1)$} & {$3.7(-5)$} & {$2.8(2)$} \\
{$5$} & {$0.1$} & {$10^{6}$} & {$0$} & {$4.3(-4)$} & {$5.8(-3)$} & {$0.51$} & {$0.24$} & {$5.4(-1)$} & {$1.9(-5)$} & {$4.6$} \\
{$5$} & {$1$} & {$10^{4}$} & {$0$} & {$3.1(-2)$} & {$2.7(-2)$} & {$1.1$} & {$0.13$} & {$1.5(-1)$} & {$3.2(-5)$} & {$3.4(2)$} \\
{$5$} & {$1$} & {$10^{5}$} & {$0$} & {$2.7(-3)$} & {$6.7(-3)$} & {$1.4$} & {$0.27$} & {$7.4(-1)$} & {$1.6(-5)$} & {$7.7(1)$} \\
{$5$} & {$1$} & {$10^{6}$} & {$0$} & {$2.2(-4)$} & {$1.5(-3)$} & {$1.8$} & {$0.58$} & {$3.7$} & {$7.2(-6)$} & {$1.1(1)$} \\
{$5$} & {$10$} & {$10^{4}$} & {$0$} & {$1.8(-2)$} & {$1.2(-2)$} & {$2.1$} & {$0.31$} & {$9.4(-1)$} & {$1.5(-5)$} & {$4.5(2)$} \\
{$5$} & {$10$} & {$10^{5}$} & {$0$} & {$1.7(-3)$} & {$3.5(-3)$} & {$2.2$} & {$0.63$} & {$4.4$} & {$6.6(-6)$} & {$1.4(2)$} \\
{$5$} & {$10$} & {$10^{6}$} & {$0$} & {$1.5(-4)$} & {$9.5(-4)$} & {$2.4$} & {$1.3$} & {$2.1(1)$} & {$2.8(-6)$} & {$2.9(1)$} \\
{$10$} & {$0.1$} & {$10^{4}$} & {$0$} & {$6.8(-2)$} & {$2.4(-1)$} & {$0.0$} & {$0.10$} & {$1.3(-2)$} & {$6.6(-5)$} & {$8.3(2)$} \\
{$10$} & {$0.1$} & {$10^{5}$} & {$0$} & {$5.8(-3)$} & {$7.0(-2)$} & {$0.0$} & {$0.10$} & {$6.5(-2)$} & {$4.2(-5)$} & {$4.0(2)$} \\
{$10$} & {$0.1$} & {$10^{6}$} & {$0$} & {$5.7(-4)$} & {$2.2(-2)$} & {$0.0$} & {$0.16$} & {$2.9(-1)$} & {$2.6(-5)$} & {$1.3(2)$} \\
{$10$} & {$1$} & {$10^{4}$} & {$0$} & {$5.6(-2)$} & {$2.2(-1)$} & {$0.0$} & {$0.10$} & {$6.5(-2)$} & {$4.2(-5)$} & {$1.3(3)$} \\
{$10$} & {$1$} & {$10^{5}$} & {$0$} & {$5.6(-3)$} & {$7.0(-2)$} & {$0.0$} & {$0.16$} & {$2.9(-1)$} & {$2.6(-5)$} & {$7.9(2)$} \\
{$10$} & {$1$} & {$10^{6}$} & {$0$} & {$3.3(-4)$} & {$3.1(-3)$} & {$0.99$} & {$0.38$} & {$1.8$} & {$1.1(-5)$} & {$5.2(1)$} \\
{$10$} & {$10$} & {$10^{4}$} & {$0$} & {$2.7(-2)$} & {$2.1(-2)$} & {$1.4$} & {$0.20$} & {$4.7(-1)$} & {$2.1(-5)$} & {$7.7(2)$} \\
{$10$} & {$10$} & {$10^{5}$} & {$0$} & {$2.3(-3)$} & {$5.3(-3)$} & {$1.6$} & {$0.42$} & {$2.3$} & {$9.8(-6)$} & {$3.2(2)$} \\
{$10$} & {$10$} & {$10^{6}$} & {$0$} & {$2.0(-4)$} & {$1.3(-3)$} & {$2.0$} & {$0.88$} & {$1.1(1)$} & {$4.2(-6)$} & {$7.4(1)$} \\
{$1$} & {$0.1$} & {$10^{4}$} & {$1$} & {$5.6(-2)$} & {$2.2(-1)$} & {$0.0$} & {$0.11$} & {$6.6(-2)$} & {$4.1(-5)$} & {$3.3(2)$} \\
{$1$} & {$0.1$} & {$10^{5}$} & {$1$} & {$3.8(-3)$} & {$1.4(-2)$} & {$0.71$} & {$0.26$} & {$3.7(-1)$} & {$2.0(-5)$} & {$1.2(1)$} \\
{$1$} & {$0.1$} & {$10^{6}$} & {$1$} & {$2.5(-4)$} & {$1.9(-3)$} & {$1.5$} & {$0.58$} & {$2.2$} & {$8.4(-6)$} & {$5.7(-2)$} \\
{$1$} & {$1$} & {$10^{4}$} & {$1$} & {$2.5(-2)$} & {$1.9(-2)$} & {$1.5$} & {$0.29$} & {$4.9(-1)$} & {$1.8(-5)$} & {$8.1(1)$} \\
{$1$} & {$1$} & {$10^{5}$} & {$1$} & {$2.1(-3)$} & {$4.6(-3)$} & {$1.8$} & {$0.61$} & {$2.4$} & {$7.9(-6)$} & {$1.3(1)$} \\
{$1$} & {$1$} & {$10^{6}$} & {$1$} & {$1.7(-4)$} & {$1.1(-3)$} & {$2.2$} & {$1.3$} & {$7.4$} & {$3.7(-6)$} & {$1.0$} \\
{$1$} & {$10$} & {$10^{4}$} & {$1$} & {$9.4(-3)$} & {$7.9(-3)$} & {$2.3$} & {$0.78$} & {$1.5$} & {$7.8(-6)$} & {$1.1(2)$} \\
{$1$} & {$10$} & {$10^{5}$} & {$1$} & {$1.4(-3)$} & {$2.9(-3)$} & {$2.5$} & {$1.4$} & {$3.2$} & {$4.5(-6)$} & {$3.3(1)$} \\
{$1$} & {$10$} & {$10^{6}$} & {$1$} & {$1.3(-4)$} & {$7.7(-4)$} & {$2.8$} & {$2.8$} & {$8.4$} & {$2.2(-6)$} & {$4.5$} \\
{$5$} & {$0.1$} & {$10^{4}$} & {$1$} & {$6.8(-2)$} & {$2.4(-1)$} & {$0.0$} & {$0.10$} & {$2.0(-2)$} & {$5.8(-5)$} & {$7.2(2)$} \\
{$5$} & {$0.1$} & {$10^{5}$} & {$1$} & {$6.0(-3)$} & {$7.2(-2)$} & {$0.0$} & {$0.11$} & {$9.9(-2)$} & {$3.7(-5)$} & {$2.8(2)$} \\
{$5$} & {$0.1$} & {$10^{6}$} & {$1$} & {$5.7(-4)$} & {$2.2(-2)$} & {$0.0$} & {$0.22$} & {$4.5(-1)$} & {$2.1(-5)$} & {$8.2(1)$} \\
{$5$} & {$1$} & {$10^{4}$} & {$1$} & {$5.7(-2)$} & {$2.2(-1)$} & {$0.0$} & {$0.11$} & {$1.0(-1)$} & {$3.7(-5)$} & {$1.2(3)$} \\
{$5$} & {$1$} & {$10^{5}$} & {$1$} & {$5.6(-3)$} & {$7.0(-2)$} & {$0.0$} & {$0.22$} & {$4.5(-1)$} & {$2.1(-5)$} & {$6.8(2)$} \\
{$5$} & {$1$} & {$10^{6}$} & {$1$} & {$4.3(-4)$} & {$6.1(-3)$} & {$0.48$} & {$0.47$} & {$2.4$} & {$9.1(-6)$} & {$7.0(1)$} \\
{$5$} & {$10$} & {$10^{4}$} & {$1$} & {$3.0(-2)$} & {$2.6(-2)$} & {$1.1$} & {$0.26$} & {$6.8(-1)$} & {$1.7(-5)$} & {$7.5(2)$} \\
{$5$} & {$10$} & {$10^{5}$} & {$1$} & {$2.6(-3)$} & {$6.6(-3)$} & {$1.4$} & {$0.55$} & {$3.3$} & {$7.7(-6)$} & {$2.6(2)$} \\
{$5$} & {$10$} & {$10^{6}$} & {$1$} & {$2.2(-4)$} & {$1.5(-3)$} & {$1.8$} & {$1.2$} & {$1.7(1)$} & {$3.2(-6)$} & {$5.3(1)$} \\
{$10$} & {$0.1$} & {$10^{4}$} & {$1$} & {$6.8(-2)$} & {$2.4(-1)$} & {$0.0$} & {$0.10$} & {$1.3(-2)$} & {$6.6(-5)$} & {$8.3(2)$} \\
{$10$} & {$0.1$} & {$10^{5}$} & {$1$} & {$1.8(-2)$} & {$1.2(-1)$} & {$0.0$} & {$0.10$} & {$3.1(-2)$} & {$5.1(-5)$} & {$5.8(2)$} \\
{$10$} & {$0.1$} & {$10^{6}$} & {$1$} & {$6.0(-4)$} & {$2.3(-2)$} & {$0.0$} & {$0.16$} & {$2.8(-1)$} & {$2.6(-5)$} & {$1.4(2)$} \\
{$10$} & {$1$} & {$10^{4}$} & {$1$} & {$6.8(-2)$} & {$2.4(-1)$} & {$0.0$} & {$0.10$} & {$5.8(-2)$} & {$4.3(-5)$} & {$1.4(3)$} \\
{$10$} & {$1$} & {$10^{5}$} & {$1$} & {$5.8(-3)$} & {$7.1(-2)$} & {$0.0$} & {$0.16$} & {$2.9(-1)$} & {$2.6(-5)$} & {$7.9(2)$} \\
{$10$} & {$1$} & {$10^{6}$} & {$1$} & {$5.7(-4)$} & {$2.2(-2)$} & {$0.0$} & {$0.32$} & {$1.3$} & {$1.3(-5)$} & {$3.9(2)$} \\
{$10$} & {$10$} & {$10^{4}$} & {$1$} & {$5.6(-2)$} & {$2.2(-1)$} & {$0.0$} & {$0.16$} & {$2.9(-1)$} & {$2.6(-5)$} & {$2.1(3)$} \\
{$10$} & {$10$} & {$10^{5}$} & {$1$} & {$4.1(-3)$} & {$1.7(-2)$} & {$0.58$} & {$0.35$} & {$1.6$} & {$1.2(-5)$} & {$6.8(2)$} \\
{$10$} & {$10$} & {$10^{6}$} & {$1$} & {$3.1(-4)$} & {$2.7(-3)$} & {$1.1$} & {$0.77$} & {$8.4$} & {$5.0(-6)$} & {$1.7(2)$}
  \enddata
  \tablecomments{The parameters $a$, $f_\Sigma$, $\beta_{z0}$, $f_\mathrm{sc}$ are the disk (input) parameters (\Tb{model-pars}). Output parameters are: turbulence strength at midplane ($\alpha_T$); the effective strength of the density fluctuations ($f_\mathrm{\delta\rho}$); the width of the dead zone ($H_\mathrm{res0}$). The dead zone occasionally disappears. The dust size distribution is characterized by the dust radius $s_D$ and the fragment radius $s_F$ and the surface area-equivalent abundance in 0.1 $\mu$m radius grains ($Z_\mathrm{0.1,eqv}$). The critical size at which bodies enter runway growth, $s_\mathrm{run}$, is given in the last column. Values written as $a(b)$ denote $a\times10^{b}$. }
\end{deluxetable*}
Assuming damping by gas drag,
\begin{equation}
  \label{eq:s-esc-def}
  f^2_{\delta\rho} q_\mathrm{gas}^2 \Omega 
  = \frac{2e^2}{T_\mathrm{drag}(e,s_\mathrm{run})}
  \quad \textrm{with}\quad
  e^2 = \frac{8\pi G_N\rho_\bullet}{3} s_\mathrm{run}^2
\end{equation}
For large bodies the drag constant $C_D=0.44$ \citep{Whipple1972} and the gas-planetesimal velocity is $v_\mathrm{gas} \approx ev_K$.  Inserting \eq{Tdrag} into \eq{s-esc-def} and solving for $s_\mathrm{run}$ gives:
\begin{eqnarray}
  \label{eq:s-run}
  s_\mathrm{run}
  &=& \left[ \frac{3^{1/2}}{4\pi C_D} \frac{H \Sigma_\mathrm{gas} a^{3/2}}{(\rho_\bullet M_\star)^{1/2}} \right]^{1/2} f_{\delta\rho} \\
  \nonumber
  &\approx& 160\ \mathrm{km} \left( \frac{H}{0.25\ \mathrm{AU}} \right)^{1/2} \left( \frac{\rho_\bullet}{\mathrm{g\ cm}^{-3}} \right)^{-1/4} \left( \frac{f_{\delta\rho}}{0.01} \right) f_\Sigma^{1/2}
\end{eqnarray}
where we assumed a solar-mass star and \eq{sig-mmsn} for $\Sigma_\mathrm{gas}(a)$. 

For the parameters corresponding to \fg{default} we took $\rho_{\bullet,P} = 2$ and found $f_{\delta\rho}=3.7\times10^{-3}$ and $s_\mathrm{run}\approx 50$ km (a bit larger than the $\approx$40 km found from \fg{default} because of the neglect of the headwind term in $v_\mathrm{gas}$ when deriving \eq{s-run}). This is the minimum radius at which planetesimals can trigger runaway grow.  Given the fact that $f_{\delta\rho}$ is only $\approx$$1/30$ of its ideal-MRI value, it is clear that a large dead zone is a necessary ingredient to expedite planet formation. 

\section{Results}
\label{sec:results}
We have conducted a parameter study, varying the strength of the magnetic field (here represented by the plasma beta parameter of the unperturbed disk, $\beta_{z0}$), the disk radius $a$, the disk mass (in terms of the MMSN, $f_\Sigma$), the ionization rate (in terms of the $f_\mathrm{sc}$ parameter; see \se{disk}), and the amount of the solids in the dust component, \ie\ in particles of radii $s\le s_F$, as given by their abundance $Z_\mathrm{dust}$, see \Tb{model-pars}. The adopted range in these parameters reflect the uncertainty regarding the physical conditions of protoplanetary disks with the default parameters (highlighted in \Tb{model-pars}) usually the central value. Since particles of radius $s_F$ dominate the \textit{dust} surface density, $\Sigma_F \approx Z_\mathrm{dust}\Sigma_\mathrm{gas}$. We generally assume that planetesimals dominate the solid surface densities, \ie\ $\Sigma_F \ll \Sigma_P = \Sigma_\mathrm{solids}$, although we will also run models where all the solids are dust ($Z_\mathrm{dust}=10^{-2}$; the gas-to-dust ratio is always fixed at 1:100).

Except for these parameters, our model is entirely self-consistent and provides: the fragmentation barrier size $s_F$, the strength of the turbulent density fluctuations $f_{\delta\rho}$ (which determines $\Delta v_{\delta\rho}$), and the magnitude of the turbulent gas velocity $\delta v_\mathrm{turb}$, which determines $\Delta v_\mathrm{TI}$. Due to the recipe-nature of the models, the computations are very fast: the parameter study is completed in a few seconds on a modern desktop PC. The output of our model in the case of a dust fraction of $Z_\mathrm{dust}=10^{-3}$ are listed in \Tb{output}. We describe some of these results in more detail below.

\subsection{The effects of dust coagulation}
\begin{figure}[tb]
  \centering
  \includegraphics[width=85mm]{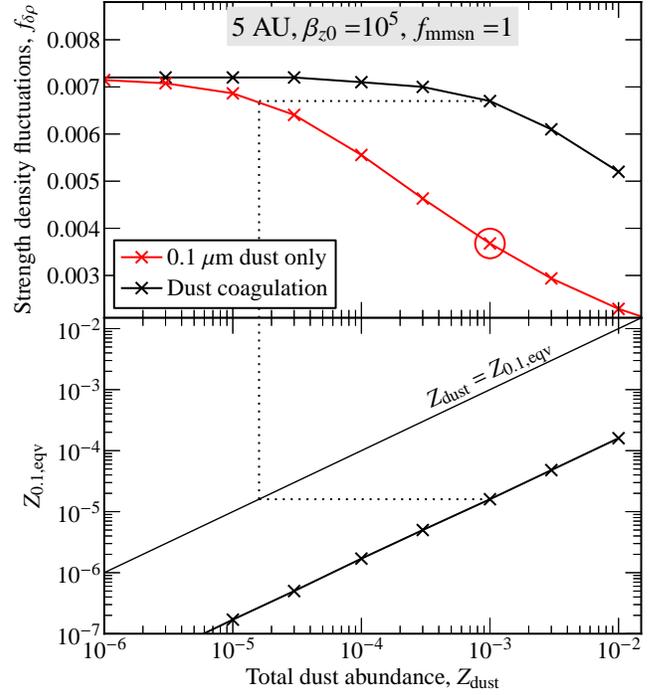}
  \caption{Top: Effective strength of the density fluctuations $f_{\delta\rho}$ as function of dust abundance for the monodisperse, 0.1 \micr\ grains (red curve) or the distribution (black curve). The red circle indicates the model corresponding to \fg{default}. Bottom: the relation between the total amount of the dust in the small particle distribution, $Z_\mathrm{dust}$, and the surface area-equivalent abundance in 0.1 \micr\ grains $Z_\mathrm{0.1,eqv}$ for the distribution case.  The dotted auxiliary illustrates that a dust fraction of $Z_\mathrm{dust}=10^{-3}$ in a distribution is equivalent to a $Z_\mathrm{0.1}\approx2\times10^{-5}$ abundance in 0.1\ \micr\ size grains.}
  \label{fig:distribution}
\end{figure}
\begin{figure}[t]
  \includegraphics[width=85mm]{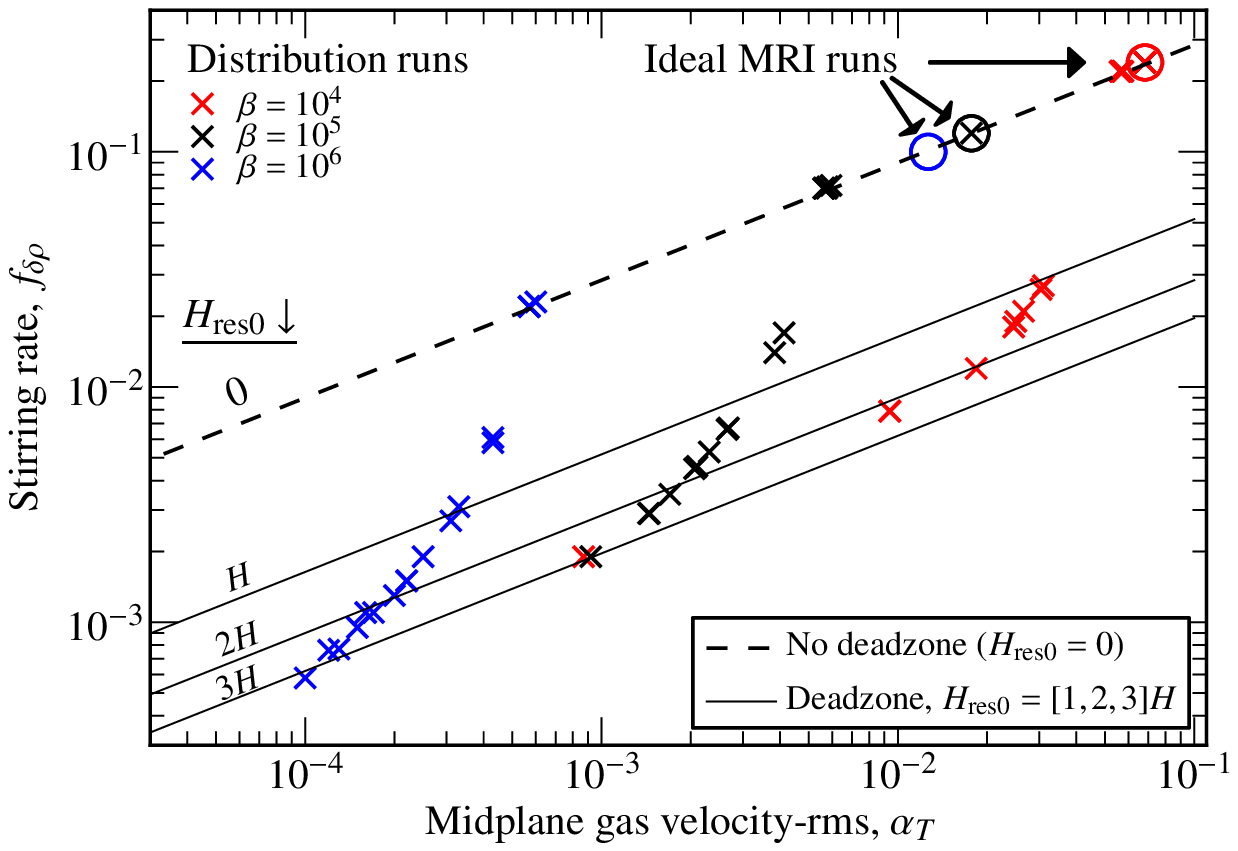}
  \caption{Scatter plot of the midplane gas-rms velocity (expressed in terms of an $\alpha$ parameter: $\alpha_T = (\delta v_\mathrm{mid}/c_s)^2$) and the dimensionless stirring parameter $f_{\delta\rho}$ for \textit{all} runs where $Z_\mathrm{dust}=10^{-3}$ (crosses). Runs are only identified by their value of $\beta_{z0}$ (colors). Ideal-MRI runs are indicated by large open circles. Lines denote the relation between $f_{\delta\rho}$ and $\alpha_T$ (\eq{fdelrho-aT}), for various deadzone sizes: $H_\mathrm{res0}=0$ (\ie\ no deadzone), 1, 2, and 3$H$.}
% The line labeled `Ideal-MRI' obeys the sound wave relation $\delta\rho/\rho \simeq v_\mathrm{mid}/c_s$. The value of $(\alpha_T, f_{\delta\rho})$ for ideal MRI is indicated by open circles. Runs that develop a dead zone fall below the ideal-MRI line, because $f_{\delta\rho}$ is reduced due to the shearing-out of the density fluctuations. The thin lines illustrate this effect for several dead zone sizes.}
  \label{fig:aTplot}
\end{figure}
In the top panel of \fg{distribution} the red curve plots the effective strength of the density fluctuations $f_{\delta\rho}$ for the standard parameters but assuming that all the dust resides in 0.1 \micr\ radius grains, \ie\ $Z_\mathrm{0.1} = Z_\mathrm{0.1,eqv} = Z_\mathrm{dust}$.  The run on which \fg{default} was based ($Z_{0.1}=10^{-3}$) is indicated by the open circle. As remarked, this run gave rise to density fluctuations of $f_\mathrm{\delta\rho} \approx 4\times10^{-3}$. When the dust abundance increases, the disk provides a larger resistivity, reducing the strength of the density fluctuations. On the other hand, when $Z_{0.1}$ is reduced, the resistivity decreases, and $f_{\delta\rho}$ increases. By $Z_{0.1}\simeq10^{-6}$ this increase has reached a saturation level. The MRI-turbulence does not become ideal, however; it turns out that the ionization rate and the field strength are too low.  Rather, the resistivity is determined through gas-phase recombination, independent of the amount of dust, and the disk still harbors a dead zone.

%; first slowly but then more pronounced until $Z\lesssim3\times10^{-4}$, where the dead zone disappears and $f_{\delta\rho}$ saturates at $f_{\delta\rho}=0.07$.  
\begin{figure*}[t]
  \centering
  \includegraphics[width=\textwidth]{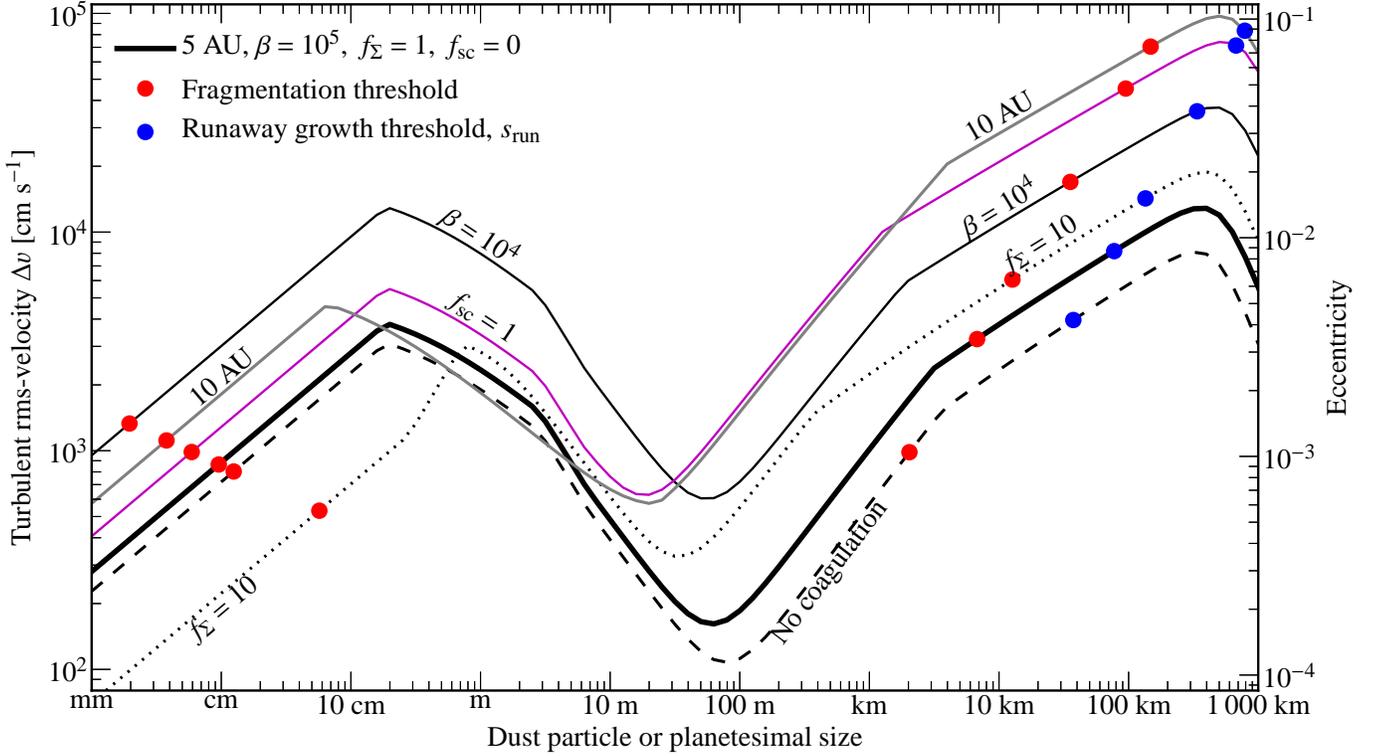}
  \caption{Effects of parameters on the equilibrium velocity as function of size. The thick black line corresponds to the default model (5 AU, $f_\Sigma=1$, $\beta_{z0}=10^5$, $f_\mathrm{sc}=0$, $Z_\mathrm{dust}=10^{-3}$) and includes the effects of dust coagulation. Other curves correspond to changing one of the parameters: no dust coagulation (dashed curve as in \fg{default}); surface density (dotted curve); magnetic field (solid black curve); solar corona (magenta curve); and disk radius (gray curve).}
  \label{fig:family}
\end{figure*}
%As seen in \fg{distribution}, the emergence of a dead zone at $Z_{0.1}\gtrsim3\times10^{-4}$ quickly causes $f_{\delta\rho}$ to decrease. This sharp transition is due to two factors. Firstly, the turbulent activity is lower in a dead zone. In addition, the turbulence-induced density fluctuations now become less effective as the associated density waves have to travel over the height of the dead zone before they reach the midplane. This shearing out of the density waves further decreases $f_\mathrm{\delta\rho}$. 
\Fg{distribution} also shows the strength of the fluctuations in case of a size distribution (black line in the top panel).  It turns out that when we account for coagulation effects, the saturation level for $f_{\delta\rho}$ ($7\times10^{-3}$) persist to a much larger $Z_\mathrm{dust}$ than in the $0.1$\ \micr\ case. We can understand this behavior from the surface area-equivalent abundance in 0.1 \micr\ grains, $Z_\mathrm{0.1,eqv}$ (\eq{sig0eff}). In the bottom panel of \fg{distribution} the relation between the total amount of dust ($Z_\mathrm{dust}$) and $Z_\mathrm{0.1,eqv}$ is shown by the black thick line. If we consider for example $Z_\mathrm{dust}=10^{-3}$ the surface area-equivalent abundance only amounts to $Z_\mathrm{0.1,eqv}\simeq10^{-5}$. The dotted lines in \fg{distribution} illustrates how to connect the result from the dust distribution model to the monodisperse models via $Z_\mathrm{0.1,eqv}$.  Thus, even for the maximum dust abundance of $Z_\mathrm{dust}\approx 10^{-2}$ one observes that the 0.1 \micr\ dust-equivalent abundance barely exceeds $10^{-4}$.  Clearly, the range in $f_{\delta\rho}$ that can be achieved for the coagulation case is less than for the monodisperse case. Accounting for dust coagulation thus weakens the dependence on $Z_\mathrm{dust}$ -- a somewhat paradoxical conclusion.  

Although \fg{distribution} represents a particular result based on (rather arbitrary) choices of the parameters, we find that this insensitivity of $f_\mathrm{\delta\rho}$ to $Z_\mathrm{dust}$ by virtue of coagulation is a robust result. \Fg{distribution} and \Tb{output} show that the depletion factors $Z_\mathrm{0.1,eqv}/Z_\mathrm{dust}$ are typically $10^{-3}$--$10^{-2}$. Interestingly, these values are in good agreement with mid-IR modeling of disk atmospheres of T-Tauri stars \citep{FurlanEtal2005,LiuEtal2012} where the depletion is sometimes interpreted as evidence for dust settling. In the context of our model, however, these depletion factors result from a competition between dust coagulation and fragmentation \citep[\cf][]{BirnstielEtal2009}.

\begin{figure*}[tbp]
  \centering
  \includegraphics[width=\textwidth]{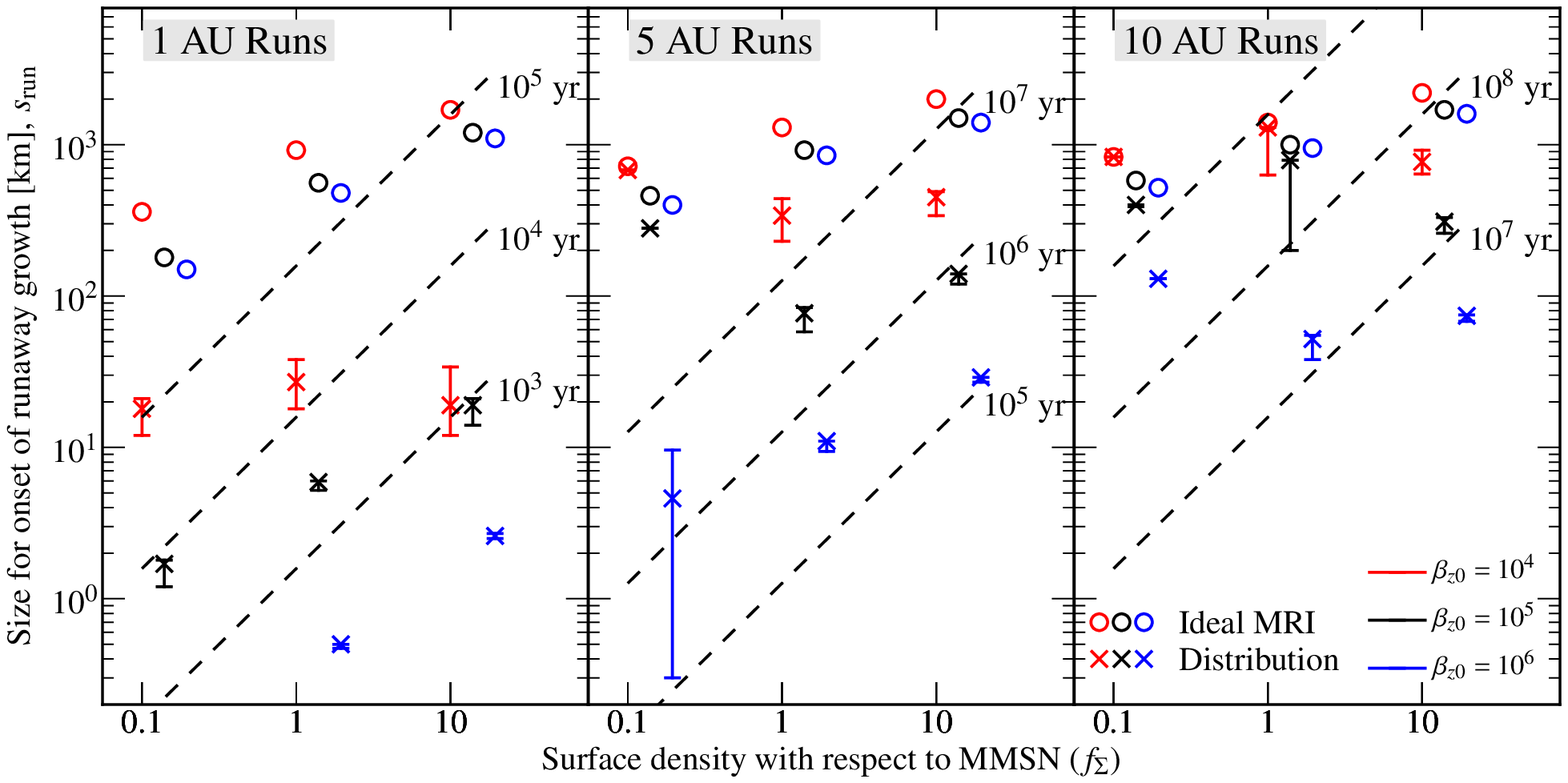}
  \caption{Minimum planetesimal size required to trigger runaway growth ($s_\mathrm{run}$).  Open circles give $s_\mathrm{run}$ for ideal MRI-turbulence conditions, whereas error bars give the range in $s_\mathrm{run}$ for the coagulation models with the dust abundance $Z_\mathrm{dust}$ ranging from 0 (no dust) to $10^{-2}$ (all the solid mass is in the dust). The runaway-growth threshold corresponding to $Z_\mathrm{dust}=10^{-3}$ is denoted by a cross. Results are shown for different disk radii (panels: 1 [left], 5 [center], and 10 AU [right]), surface density in terms of the MMSN ($x$-axis), and strength of the vertical net field (colors, slightly offset).  Note that dust fractions of 0\% do not necessarily imply ideal MRI conditions (because of gas-phase recombination) and 100\% dust fraction are not necessarily imply a large abundance of very small grains (because of their coagulation). Dashed lines correspond to the collision timescale among the planetesimal population without gravitational focusing.}
  \label{fig:s-rg}
\end{figure*}
\begin{figure*}[tbp]
  \centering
  \includegraphics[width=\textwidth]{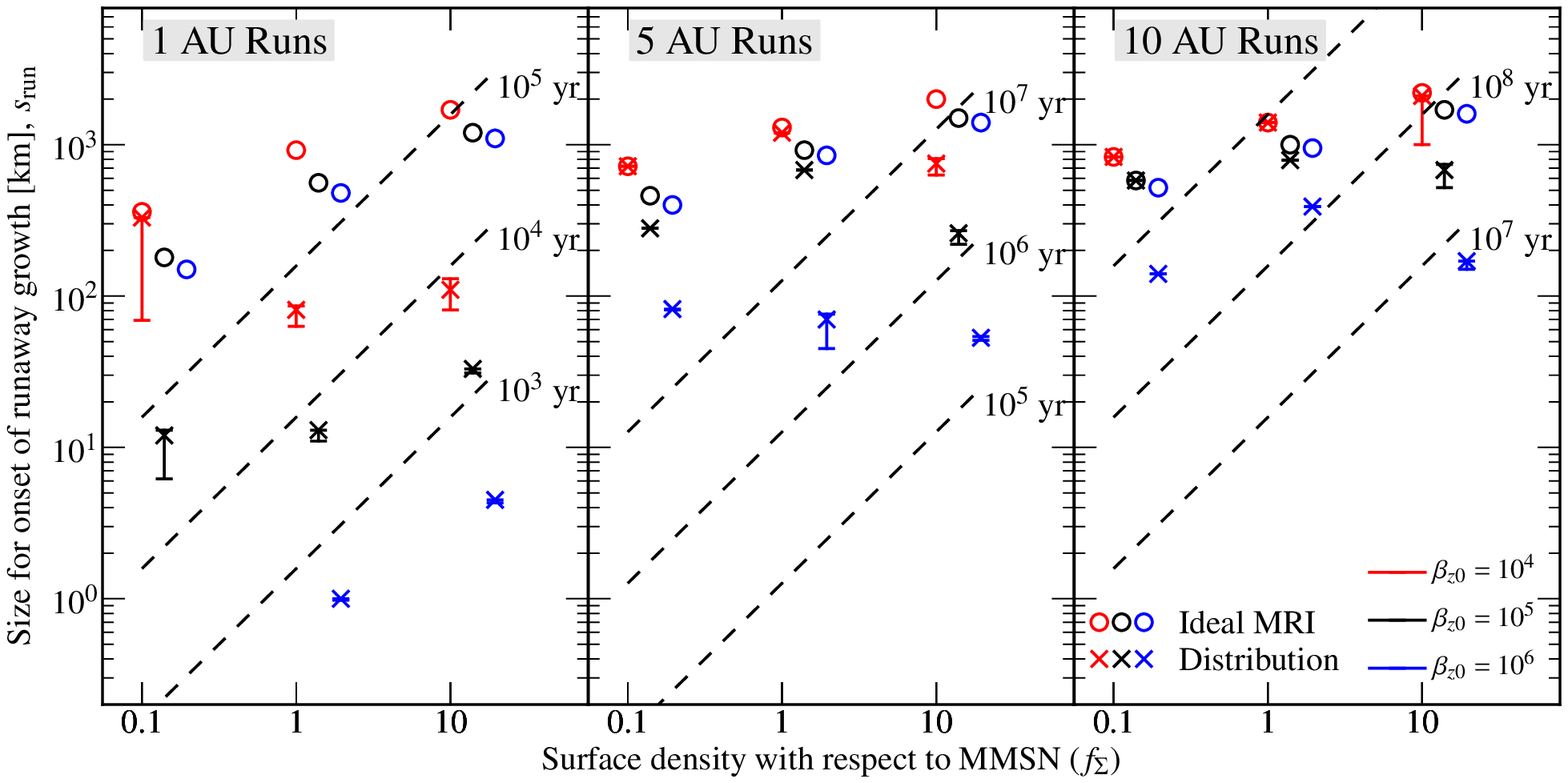}
  \caption{Same as \fg{s-rg} but in case with an active solar corona ($f_\mathrm{sc}=1$), \ie\ at ionization levels much higher than in \fg{s-rg}.}
  \label{fig:s-rg2}
\end{figure*}

% , which is still too low to play any role in the ionization balance of the gas 
%any $Z_\mathrm{dust}\le10^{-2}$ will give the same $f_{\delta\rho}$. 

\subsection{The relation between turbulent-$\alpha$ and turbulent stirring}
\Fg{aTplot} is a scatter plot for $\alpha_T$, a proxy for the rms-gas velocities at the midplane, and $f_{\delta\rho}$ for all runs of the parameter study where $Z_\mathrm{dust}=10^{-3}$ (crosses) . Thus, while $\alpha_T$ determines the relative turbulent velocity among small dust particles, $f_{\delta\rho}$ provides that among big bodies. Using the definition of $\delta v_\mathrm{mid}$ (\eq{v-mid}) and $f_{\delta\rho}$ (\eq{e-tidf}) we obtain the relation
\begin{equation}
  f_{\delta\rho} \approx 0.9 \frac{\alpha_T^{1/2}}{1 +H_\mathrm{res0}/H}.
  \label{eq:fdelrho-aT}
\end{equation}
%The corresponding values of $f_\Sigma$ and $a_0$ can be found by inspection with \Tb{output}.

In \fg{aTplot} we distinguish runs of different $\beta_{z0}$ by color and also show the results in the ideal-MRI limit (open circles).  Models where the MRI is ideal obey $f_{\delta\rho}\approx0.9\alpha_T^{1/2}$, which is indicated by the dashed line in \fg{aTplot}. In ideal MRI the stirring rate (and $\alpha_T$) are not very sensitive to the value of the external field, $\beta_{z0}$; an asymptotic limit of $\alpha_T\approx10^{-2}$ is reached when $\beta_{z0}\rightarrow\infty$ \citep[OH11]{DavisEtal2010,SuzukiEtal2010}. When the ideal-MRI assumption is relaxed, the level of turbulence activity ($\alpha_T$) can become much lower. The precise level now depends on the disk parameters ($\Sigma_\mathrm{gas}$, radius, ionization sources), the dust content, and (most importantly) the value of the external field. However, as long as a dead zone does not develop, runs still obey the ideal-MRI relation for $\alpha_T$ and $f_{\delta\rho}$ (dashed line). This means that $f_{\delta\rho}$ remains rather large, yielding a large threshold for runaway growth (see \eqp{s-run}) even when the disk has become quite laminar. For a more significant reduction in $f_{\delta\rho}$ a dead zone is a prerequisite, because the shearing-out effect distorts the density fluctuations.  A strong deadzone will decrease $f_{\delta\rho}$ by an order of magnitude -- equivalent to a reduction of $\alpha_T$ by a factor of 100. These two effects are both controled by the plasma-beta parameter: a lower external field decreases $\alpha_T$ and increases the likelihood of a dead zone.

\subsection{Effects of parameter variation on turbulent velocities}
\Fg{family} shows the sensitivity of the equilibrium velocity $\Delta v$ against varying the input parameters. In \fg{family} the dashed line is the same as \fg{default} (\ie\ no dust coagulation) and the thick line represents our default model (\ie\ with dust coagulation). Clearly, coagulation, which reduces the effective dust abundance, increases the equilibrium velocity. But the effect is rather modest due to gas-phase recombination combined with a relatively low ionization flux; a dead zone will exist even in the absence of dust.

Each other curve in \fg{family} reflects a change in one parameter with respect to the default model. Increasing the surface density, increases slightly the amount of turbulent excitation as the strength of the density fluctuations scale with the gas density (via the $q_\mathrm{gas}$ term in \eq{De-def}). Another (minor) influence is that the higher density suppresses turbulent velocities for small particles. Consequently, $s_F$ is larger, coagulation more efficient, and the resistivity (due to dust) decreases. On the other hand stronger gas damping and a larger dead zone due to a larger column will reduce $\Delta v$. For the excitation rate of planetesimals, it turns out that these effects cancel each other to a large extent.

Another way to increase the turbulent stirring is to increase the strength of the magnetic field, in \fg{family} represented by a decrease in the plasma beta parameter to $\beta_{z0}=10^4$. A stronger field greatly enhances the turbulent activity in both dead zone, which is somewhat smaller but still present, and the active layers. The dead zone disappears, however, if instead the ionization flux is greatly increased, as we hypothesized by including the large contribution from the stellar corona ($f_\mathrm{sc}=1$; purple curve). As the dead zone disappears, the turbulence-induced density fluctuations penetrate the midplane directly (\ie\ they do not suffer from the shear-out effect) and the excitation of planetesimals becomes much stronger. For $\sim$100 km bodies, turbulence stirring reaches eccentricities close to $\sim$0.1.

In the outer regions of the disk, the dead zone tends to be absent as the column density is lower. By 10 AU (gray line) the dead zone has disappeared.  In the outer planetary systems, turbulent stirring is thus expected to be much more violent as compared to the inner regions.  

\subsection{The threshold for runaway growth}
The prerequisite for a system of planetesimals to initiate runaway accretion is that their escape velocities exceed their random motions, $v_\mathrm{esc}>\Delta v$, which enhances the collisional cross section by a factor $(v_\mathrm{esc}/v)^2$ over the geometrical cross section $\sim$$\pi s_P^2$. As reviewed in \se{intro}, during runaway growth gravitational focusing ensures a positive feedback and the ensuing runaway growth will persist. 

In \eq{s-run} the size $s_\mathrm{run}$ corresponding to the point where $v_\mathrm{esc}=\Delta v_{\delta\rho}$ was derived assuming gas damping and a quadratic drag law. In \fg{s-rg} the numerically-derived $s_\mathrm{run}$, which includes tidal damping, is plotted as function of the parameters that we have investigated in this paper: the strength of the net vertical magnetic field (colors); the disk radius (panels), the disk mass $f_\Sigma$ ($x$-axis). However, only runs with $f_\mathrm{sc}=0$ are shown.  The runaway growth size resulting from ideal MRI conditions are indicated by open circles and error bars show $s_\mathrm{run}$ for the resistive case with a dust distribution. The crosses correspond to a dust fraction of $10^{-3}$. Runs without dust ($Z_\mathrm{dust}=0$) are also included but in most cases the gas layer is still sufficiently thick to prevent the MRI from becoming ideal. 

\Fg{s-rg} illustrates once again that dust coagulation tends to render the results irrelevant to the dust content: in many of the runs the error bars are virtually absent. The large error bar corresponding to the 5 AU, $f_\Sigma=0.1$, $\beta_{z0}=10^{6}$ runs is an exception. Investigation showed that  for these parameters the turbulent velocity line $\Delta v_{\delta\rho}(s)$ lies very close to the escape velocity curve, $v_\mathrm{esc}$ -- in a way much similar as $\Delta v_{\delta\rho}$ and the weak material strength curve of \fg{default} coincide. Consequently, a small change in the dust properties, resulting in a small translation of $\Delta v_{\delta\rho}$ with respect to $v_\mathrm{esc}$, gives rise to a large change in the intersection point of these curves ($s_\mathrm{run}$).

As was found before, the minimum RG-radius $s_\mathrm{run}$ is also rather insensitive to the disk mass. A more massive gas disk may cause the dead zone to increase somewhat, reducing the turbulent activity in the midplane ($f_\mathrm{\delta\rho}$). On the other hand, an increased gas density results in stronger fluctuations, \ie\ a larger torque. As mentioned, these effects tend to cancel each other to a large extent.

More important is the strength of the vertical net field. There is a clear and positive correlation between $B_{z0}$ (or $\beta_{z0}$): the stronger the field, the stronger the turbulence, and the larger $s_\mathrm{run}$. The importance of the strength of the net vertical magnetic field was already highlighted by \citet{OkuzumiHirose2012}. In essence the result follows from the observed (empirical) correlation between the stresses the MRI attains in the saturated state and the value of $B_{z0}$ (see OH11 and Paper I).

In \fg{s-rg2} the same plot is shown, but then for the runs that include the high ionization levels resulting from the solar corona, $f_\mathrm{sc}=1$. Because of the much higher flux $s_\mathrm{run}$ increases, sometimes significantly in cases that the dead zone has disappeared. Apart from this, the general trends (insensitivity to $Z_\mathrm{dust}$ and $f_\Sigma$ and a stronger dependence on $\beta_{z0}$) are still apparent.

\section{Discussion}
\label{sec:discuss}
\subsection{Implications for planetesimal formation and accretion}
Little is still known on how the formation of planetesimals proceeds.  Coagulation to the km-size regime by incremental accretion is hindered by several `barriers', which all find their root in the increase of relative velocities as particle sizes approach $T_\mathrm{drag}\Omega\approx1$. This corresponds to the peak of the turbulent inertia regime, where $\Delta v_\mathrm{TI} \approx \delta v_\mathrm{mid}$ (see \fg{default}). However, particles will start to fragment much earlier, possibly already at velocities lower than $1\ \mathrm{m\ s}^{-1}$ \citep{BeitzEtal2011}. In the ice-dominated outer disk, the threshold is expected to be much larger though, perhaps $50\ \mathrm{m\ s}^{-1}$ \citep{WadaEtal2009}.  At 5 AU, such a fragmentation threshold corresponds to midplane density fluctuations of $\delta\rho/\rho \approx v_\mathrm{frag}/c_s \approx 0.07$.

In the absence of a dead zone this is also the value of $f_{\delta\rho}$, which implies that the runaway grow radius $s_\mathrm{run}$ is large, approaching $10^3$ km (\eq{s-run}). Such a large threshold size for planetesimals is problematic, however, because of the long collision timescale. \textit{Without} gravitational focusing the collision timescale between two similar-size bodies reads
\begin{eqnarray}
  \label{eq:Tcol}
  T_\mathrm{col} 
  &\sim& \frac{s_P \rho_{\bullet,P}}{\Sigma_\mathrm{solid}} \Omega^{-1} \\ \nonumber
  &\sim& 10^7\ \mathrm{yr}\ f_\Sigma^{-1} \left( \frac{s_P}{100\ \mathrm{km}} \right) \left( \frac{a}{\mathrm{5\ AU}} \right)^3 \left( \frac{\rho_{\bullet,P}}{2\ \mathrm{g\ cm}^{-3}} \right),
\end{eqnarray}
which will rival the lifetime of the nebula $T_\mathrm{neb}$ (several $10^6$ yr).  Large planetesimals (small embryos) of $s<s_\mathrm{run}$ may thus see the nebula dissipating away long before they have reached the critical size at which they would have been able to bind the gas.  Altogether these considerations imply that the runway grow barrier at $s_\mathrm{run}$ is at least as formidable a bottleneck to planet formation than the fragmentation barrier at $T_\mathrm{drag}\Omega=1$.  To further illustrate this point, we have drawn isocontours of $T_\mathrm{col}$ in \fg{s-rg}. Thus, runs whose points lie much above the $10^{6}$ yr contour may never experience a classical runaway growth phase.  At 1 AU the timescales are not problematic: growth timescales are sufficiently short even in the (unlikely) case that the MRI is ideal.  By 5 AU the timescales already become uncomfortably long: the strength of the field has to subside to levels corresponding to $\beta_{z0}>10^5$. Beyond 10 AU, where it becomes ever-harder to preserve a dead zone, these results imply that the disk should become laminar for runaway growth to commence. 

The problem is that planetesimal self-coagulation at sizes below $s_\mathrm{run}$ in the absence of gravitational focusing is slow: growth timescales increase with the cube of the disk radius (\eq{Tcol}). If gravitational focusing would be initiated, it tends to mitigate the dependence on disk radius; that is, gravitational focusing factors increase with increasing $a$ \citep{Rafikov2006}. Nevertheless, from a timescale perspective small planetesimals are often preferred \citep[\eg][]{KenyonBromley2009,FortierEtal2013}. Thus, the timescale problem, already problematic in the classical models, is exacerbated if planetesimals are required to have a minimum size corresponding to $s_\mathrm{run}$. Formation of massive cores by accretion of planetesimals seems impossible in the outer regions of turbulent disks.

The assumption in \eq{Tcol} is that the inclination of planetesimals ($i$) are similar to their eccentricities ($i\approx e/2$). Recently, \citet{YangEtal2012} measured the inclination stirring in ideal MRI simulations and found some evidence that the stirring is anisotropic; they found $i\approx e/5$. If this also holds for (resistive) MRI turbulence, the collision timescale \eq{Tcol} will be lower by a similar factor, because the bodies are more densely populated near the midplane, which alleviates the timescale issue to some degree. We encourage further investigation into the anisotropy of planetesimal motions caused by turbulent excitation.

\subsection{Revival of the classical planet formation scenario?}
Several caveats in the above reasoning could revive the desired setting for planet formation, \ie\ a situation where big embryos accrete smaller bodies at large focusing factors. Generally, this can be done in two ways: either by invoking a mechanisms that produces a few large planetesimals seeds or by decreasing $s_\mathrm{run}$. The latter is the most obvious route and implies that the magnetic field, $B_\mathrm{z0}$, must decrease as the effect of other parameters is relatively minor (see \fg{s-rg}). For example, when $\beta_{z0}>10^6$ the turbulent activity at 5 AU has decreased to levels where $s_\mathrm{run}\sim10$ km. The question thus becomes on which timescales the net vertical field will decay, \ie\ when the disk becomes laminar. The effects of ambipolar diffusion (not included here) may accelerate the transition to a laminar disk \citep{BaiStone2013}.

Alternatively, the $s_\mathrm{run}$ barrier can be overcome by relaxing the assumption that all planetesimals are of the same size or that their collisions occur at the same (relative) velocity. Planetesimals could be formed with a wide range of sizes; formation of a few $\sim$$10^3$ km embryo seeds among a sea of smaller bodies would readily lead to large focusing factors. Stochasticity in the velocity distribution and in the collision outcomes may offer pathways to broaden the size distribution \citep{WindmarkEtal2012i,GaraudEtal2013}. When the planetesimal (initial) mass function happens to obey the right properties -- some big, most small -- a conducive environment for growth is present.

A more direct way to envision the (sudden) emergence of large embryos is through outward scattering or migration of seeds from the inner solar system. Scattering has been observed in several core formation studies \citep{WeidenschillingEtal1997,ThommesEtal2008}. In the context of this work one needs to scatter a body of $s>s_\mathrm{run}$ as otherwise this body will have to growth via slow coagulation without focusing. Secondly, for strong scattering, the escape velocity of the scatterer must be comparable to the local Keplerian velocity as otherwise bodies cannot escape the potential well. Strong scattering events are therefore more difficult to achieve in the very inner planetary system. Altogether, the parameter space for outward scattering may be limited. Type I migration could be directed outwards (usually it is directed towards the star) if certain thermodynamic requirements of the disks are met \citep{PaardekooperMellema2006}. Planetesimal-driven \citep{CapobiancoEtal2011,OrmelEtal2012} or turbulence-driven migration (\citealt{Nelson2005}; Paper I) are other mechanisms which would invalidate the local picture.

\subsection{Delayed runaway growth?}
Studies addressing the runaway growth stage often assume laminar conditions in which runaway growth takes off instantaneously.  There is no source of external excitation; planetesimals are only stirred by the planetary embryos. Even then it is difficult to form big cores within $\sim$Myr when the planetesimal radius $s_P$ is large \citep{LevisonEtal2010,OrmelEtal2010i,KenyonBromley2010,FortierEtal2013}. For small planetesimal sizes self-fragmentation and radial orbital decay is also a concern \citep{KobayashiEtal2010,KobayashiEtal2011}.  

Suppose that initially $s_\mathrm{run}$ is large, but that eventually the planetesimals will breach this barrier because of self-coagulation or the decay of the MRI turbulence. This would delay the onset of runaway growth. Such a scenario has been suggested by \citet{GresselEtal2011} and was also found in one of the runs conducted in \citet{OrmelEtal2010i}. Turbulence-delayed runaway growth differs from the classical (laminar) models, because of the initial insignificance of viscous stirring by embryos. Viscous stirring imposes a negative feedback to the growth: stirring rates increase during the growth of embryos.  However, turbulent stirring is independent of the embryo mass. As a result, embryos emerge quickly from the bodies that first breach $s>s_\mathrm{run}$. Another important example of an external stirring mechanism is the secular forcing in binary systems \citep[\eg][]{Meschiari2012i,PaardekooperEtal2012}. Delayed runaway growth scenarios that are dominated by external stirring mechanisms are worth further investigation.

\subsection{Scenarios involving small particles}
Alternatively, one can envision that the first generation of planetesimals grew larger by sweeping up smaller particles, simply by virtue of its geometrical cross section. This idea is attractive because it is a well-attested laboratory finding that small projectiles will stick to larger bodies \citep{TeiserWurm2009}. It could be a way to form and grow early planetesimals \citep{XieEtal2010,WindmarkEtal2012}. However, because growth proceeds without focusing, situations where $s_\mathrm{run}$ is large will experience the same timescale problem.

Finally, planetesimals may form big out of a population of pre-planetesimal particles, possibly from a turbulent concentration mechanism \citep{JohansenEtal2007,CuzziEtal2010} or through streaming or other laminar instabilities \citep{YoudinGoodman2005,ShiChiang2013}. Thereafter, they can transition quickly into cores by accreting directly from the pre-planetesimal population \citep{LambrechtsJohansen2012,MorbidelliNesvorny2012}. No planetesimals are needed in this scenario, but a single massive-enough seed must be formed. The drag-enhanced gravitational focusing factors could have been very large \citep{OrmelKlahr2010,PeretsMurray-Clay2011}. Note that these studies have assumed circular orbits; but turbulence stirring may give small embryos some eccentricity (\fg{default}), until after $\approx$$10^3$ km tidal damping sets in.  

\section{Summary}
\label{sec:summary}
We have extended previous modeling of MRI-turbulence (Paper I) by including a model for the size distribution of dust grains. We assumed that the small dust population is in a coagulation/fragmentation balance where particles stick until they meet a fragmentation threshold at a radius $s_F$. By slightly modifying the prescription of \citet{BirnstielEtal2011}, we have characterized the dust size distribution in terms of two power-laws that merge at the dust radius $s_D$, below which Brownian motion efficiently removes small grains. We have expressed the dust distribution in terms of an surface area-equivalent abundance in $0.1\ \mu$m radius particles, $Z_\mathrm{0.1,eqv}$, which can be used in the calculation of the resistivity profile of the gas. The model for the dust size distribution is then combined with previously-presented recipes that provide the state of the MRI turbulence and the extent of the dead zone (\citealt{Okuzumi2009}, OH11, Paper I). By iterating these prescriptions one obtains a self-consistent description of MRI-turbulence, in which most free parameters can be eliminated. Naturally, the state of the turbulence depends on disk parameters as the net vertical magnetic field and the ionization sources. 

Our results can be used to constrain planet formation scenarios.  In the future we will include the set of recipes described in this work with a previous model for core growth \citep{OrmelKobayashi2012}, which pertains the oligarchic growth state of planet formation where planetary embryos accrete the planetesimals at large gravitational focusing factors. A key question here is how large the focusing factors are, \ie\ whether they are dominated by viscous stirring of embryos or by external stirring due to the turbulence-induced density fluctuations.

In this paper, we have focused on the planet formation phase that precedes oligarchy -- the runaway growth phase -- which is an important cornerstone of the classical planet formation model as it provides a population of planetary embryos.  However, runaway growth is only triggered when the excitation of the planetesimal population is low; a presumption that, we find, is prone to be violated in a turbulent disk. Generally, planetesimals need to exceed a threshold radius $s_\mathrm{run}$, beyond which their escape velocities are large enough to trigger runaway growth. In many cases $s_\mathrm{run}$ is rather large and the corresponding collision timescales are $\gg$Myr, much longer than the lifetime of the nebula.

Our main findings are the following:
\begin{enumerate}
  \item Coagulation causes the surface area in dust grains to decrease. We find typical depletion factor of $Z_\mathrm{0.1,eqv}/Z_\mathrm{dust}\approx10^{-3}$--$10^{-2}$, which compare favorably with mid-IR observations of T-Tauri stars. As a result, the dust abundance will little affect the properties of the turbulence; it cannot be invoked to mitigate the effects of the MRI. The development of dead zones, if they appear, is solely by virtue of gas-phase chemistry.
  \item Although generally insensitive to the dust abundance, the level of turbulent activity depends rather strongly on the value of the net vertical field, $B_{z0}$. A more laminar disk (low $B_{z0}$) offers a significantly more conducive environment for planetesimal accretion, as well as for planetesimal formation \citep{OkuzumiHirose2012}.
  \item In the inner disk regions, high densities ensure that collision timescales among planetesimals are short and that gas drag efficiently damps their eccentricities. As a result, the condition for runaway growth will be met in the inner disk. 
  \item In the outer disk (beyond 5 AU) the classical scenario for planet formation, which involves runaway growth, is incompatible with a turbulent disk. Sufficiently short accretion timescales are only achieved when the turbulent activity subsides to levels corresponding to midplane-alpha values below $\alpha_T\approx10^{-3}$ (5AU) to $10^{-4}$ (10 AU; see \fg{aTplot}).
\end{enumerate}

\acknowledgments
C.W.O.\ acknowledges support for this work by NASA through Hubble Fellowship grant \#HST-HF-51294.01-A awarded by the Space Telescope Science Institute, which is operated by the Association of Universities for Research in Astronomy, Inc., for NASA, under contract NAS 5-26555. S.O. was supported by Grant-in-Aid for JSPS Fellows (22 $\cdot$ 7006) from MEXT of Japan. The authors thank Xuening Bai, Til Birnstiel, Jeff Cuzzi, Tristian Guillot, Shigeru Ida, and the referee for comments and helpful suggestions.

\bibliographystyle{apj}
\bibliography{mybibl,ads,arXiv}

\begin{thebibliography}{87}
\expandafter\ifx\csname natexlab\endcsname\relax\def\natexlab#1{#1}\fi

\bibitem[{{{\'A}d{\'a}mkovics} {et~al.}(2011){{\'A}d{\'a}mkovics}, {Glassgold},
  \& {Meijerink}}]{AdamkovicsEtal2011}
{{\'A}d{\'a}mkovics}, M., {Glassgold}, A.~E., \& {Meijerink}, R. 2011, \apj,
  736, 143

\bibitem[{{Bai} \& {Stone}(2013)}]{BaiStone2013}
{Bai}, X.-N. \& {Stone}, J.~M. 2013, ArXiv e-prints:1301.0318

\bibitem[{{Balbus} \& {Hawley}(1991)}]{BalbusHawley1991}
{Balbus}, S.~A. \& {Hawley}, J.~F. 1991, \apj, 376, 214

\bibitem[{{Beitz} {et~al.}(2011){Beitz}, {G{\"u}ttler}, {Blum}, {Meisner},
  {Teiser}, \& {Wurm}}]{BeitzEtal2011}
{Beitz}, E., {G{\"u}ttler}, C., {Blum}, J., {Meisner}, T., {Teiser}, J., \&
  {Wurm}, G. 2011, \apj, 736, 34

\bibitem[{{Benz} \& {Asphaug}(1999)}]{BenzAsphaug1999}
{Benz}, W. \& {Asphaug}, E. 1999, Icarus, 142, 5

\bibitem[{{Birnstiel} {et~al.}(2009){Birnstiel}, {Dullemond}, \&
  {Brauer}}]{BirnstielEtal2009}
{Birnstiel}, T., {Dullemond}, C.~P., \& {Brauer}, F. 2009, \aap, 503, L5

\bibitem[{{Birnstiel} {et~al.}(2011){Birnstiel}, {Ormel}, \&
  {Dullemond}}]{BirnstielEtal2011}
{Birnstiel}, T., {Ormel}, C.~W., \& {Dullemond}, C.~P. 2011, \aap, 525, A11

\bibitem[{{Blaes} \& {Balbus}(1994)}]{BlaesBalbus1994}
{Blaes}, O.~M. \& {Balbus}, S.~A. 1994, \apj, 421, 163

\bibitem[{{Blum}(2004)}]{Blum2004}
{Blum}, J. 2004, in Astronomical Society of the Pacific Conference Series, Vol.
  309, Astrophysics of Dust, ed. A.~N. {Witt}, G.~C. {Clayton}, \& B.~T.
  {Draine}, 369

\bibitem[{{Capobianco} {et~al.}(2011){Capobianco}, {Duncan}, \&
  {Levison}}]{CapobiancoEtal2011}
{Capobianco}, C.~C., {Duncan}, M., \& {Levison}, H.~F. 2011, \icarus, 211, 819

\bibitem[{{Carballido} {et~al.}(2011){Carballido}, {Bai}, \&
  {Cuzzi}}]{CarballidoEtal2011}
{Carballido}, A., {Bai}, X.-N., \& {Cuzzi}, J.~N. 2011, \mnras, 415, 93

\bibitem[{{Chambers}(2008)}]{Chambers2008}
{Chambers}, J. 2008, Icarus, 198, 256

\bibitem[{{Chiang} \& {Laughlin}(2012)}]{ChiangLaughlin2012}
{Chiang}, E. \& {Laughlin}, G. 2012, ArXiv e-prints:1211.1673

\bibitem[{{Chokshi} {et~al.}(1993){Chokshi}, {Tielens}, \&
  {Hollenbach}}]{ChokshiEtal1993}
{Chokshi}, A., {Tielens}, A.~G.~G.~M., \& {Hollenbach}, D. 1993, \apj, 407, 806

\bibitem[{{Cuzzi} {et~al.}(2005){Cuzzi}, {Ciesla}, {Petaev}, {Krot}, {Scott},
  \& {Weidenschilling}}]{CuzziEtal2005}
{Cuzzi}, J.~N., {Ciesla}, F.~J., {Petaev}, M.~I., {Krot}, A.~N., {Scott},
  E.~R.~D., \& {Weidenschilling}, S.~J. 2005, in Astronomical Society of the
  Pacific Conference Series, Vol. 341, Chondrites and the Protoplanetary Disk,
  ed. {A.~N.~Krot, E.~R.~D.~Scott, \& B.~Reipurth}, 732--+

\bibitem[{{Cuzzi} \& {Hogan}(2003)}]{CuzziHogan2003}
{Cuzzi}, J.~N. \& {Hogan}, R.~C. 2003, \icarus, 164, 127

\bibitem[{{Cuzzi} {et~al.}(2010){Cuzzi}, {Hogan}, \& {Bottke}}]{CuzziEtal2010}
{Cuzzi}, J.~N., {Hogan}, R.~C., \& {Bottke}, W.~F. 2010, Icarus, 208, 518

\bibitem[{{Davis} {et~al.}(2010){Davis}, {Stone}, \& {Pessah}}]{DavisEtal2010}
{Davis}, S.~W., {Stone}, J.~M., \& {Pessah}, M.~E. 2010, \apj, 713, 52

\bibitem[{{Fortier} {et~al.}(2013){Fortier}, {Alibert}, {Carron}, {Benz}, \&
  {Dittkrist}}]{FortierEtal2013}
{Fortier}, A., {Alibert}, Y., {Carron}, F., {Benz}, W., \& {Dittkrist}, K.-M.
  2013, \aap, 549, A44

\bibitem[{{Furlan} {et~al.}(2005){Furlan}, {Calvet}, {D'Alessio}, {Hartmann},
  {Forrest}, {Watson}, {Uchida}, {Sargent}, {Green}, \&
  {Herter}}]{FurlanEtal2005}
{Furlan}, E., {Calvet}, N., {D'Alessio}, P., {Hartmann}, L., {Forrest}, W.~J.,
  {Watson}, D.~M., {Uchida}, K.~I., {Sargent}, B., {Green}, J.~D., \& {Herter},
  T.~L. 2005, \apjl, 628, L65

\bibitem[{{Gammie}(1996)}]{Gammie1996}
{Gammie}, C.~F. 1996, \apj, 457, 355

\bibitem[{{Garaud} {et~al.}(2013){Garaud}, {Meru}, {Galvagni}, \&
  {Olczak}}]{GaraudEtal2013}
{Garaud}, P., {Meru}, F., {Galvagni}, M., \& {Olczak}, C. 2013, \apj, 764, 146

\bibitem[{{Gressel} {et~al.}(2011){Gressel}, {Nelson}, \&
  {Turner}}]{GresselEtal2011}
{Gressel}, O., {Nelson}, R.~P., \& {Turner}, N.~J. 2011, \mnras, 415, 3291

\bibitem[{{Gressel} {et~al.}(2012){Gressel}, {Nelson}, \&
  {Turner}}]{GresselEtal2012}
---. 2012, \mnras, 422, 1140 (GNT12)

\bibitem[{{Guilloteau} {et~al.}(2012){Guilloteau}, {Dutrey}, {Wakelam},
  {Hersant}, {Semenov}, {Chapillon}, {Henning}, \&
  {Pi{\'e}tu}}]{GuilloteauEtal2012}
{Guilloteau}, S., {Dutrey}, A., {Wakelam}, V., {Hersant}, F., {Semenov}, D.,
  {Chapillon}, E., {Henning}, T., \& {Pi{\'e}tu}, V. 2012, \aap, 548, A70

\bibitem[{{Hayashi} {et~al.}(1985){Hayashi}, {Nakazawa}, \&
  {Nakagawa}}]{HayashiEtal1985}
{Hayashi}, C., {Nakazawa}, K., \& {Nakagawa}, Y. 1985, in Protostars and
  Planets II, ed. {D.~C.~Black \& M.~S.~Matthews} (Univ. of Arizona Press,
  Tuscon), 1100--1153

\bibitem[{{Heinemann} \& {Papaloizou}(2012)}]{HeinemannPapaloizou2012}
{Heinemann}, T. \& {Papaloizou}, J.~C.~B. 2012, \mnras, 419, 1085

\bibitem[{{Hughes} {et~al.}(2011){Hughes}, {Wilner}, {Andrews}, {Qi}, \&
  {Hogerheijde}}]{HughesEtal2011}
{Hughes}, A.~M., {Wilner}, D.~J., {Andrews}, S.~M., {Qi}, C., \& {Hogerheijde},
  M.~R. 2011, \apj, 727, 85

\bibitem[{{Ida} {et~al.}(2008){Ida}, {Guillot}, \& {Morbidelli}}]{IdaEtal2008}
{Ida}, S., {Guillot}, T., \& {Morbidelli}, A. 2008, \apj, 686, 1292

\bibitem[{{Ilgner} \& {Nelson}(2006)}]{IlgnerNelson2006}
{Ilgner}, M. \& {Nelson}, R.~P. 2006, \aap, 445, 205

\bibitem[{{Johansen} {et~al.}(2007){Johansen}, {Oishi}, {Low}, {Klahr},
  {Henning}, \& {Youdin}}]{JohansenEtal2007}
{Johansen}, A., {Oishi}, J.~S., {Low}, M., {Klahr}, H., {Henning}, T., \&
  {Youdin}, A. 2007, \nat, 448, 1022

\bibitem[{{Kenyon} \& {Bromley}(2009)}]{KenyonBromley2009}
{Kenyon}, S.~J. \& {Bromley}, B.~C. 2009, \apjl, 690, L140

\bibitem[{{Kenyon} \& {Bromley}(2010)}]{KenyonBromley2010}
---. 2010, \apjs, 188, 242

\bibitem[{{Kobayashi} {et~al.}(2011){Kobayashi}, {Tanaka}, \&
  {Krivov}}]{KobayashiEtal2011}
{Kobayashi}, H., {Tanaka}, H., \& {Krivov}, A.~V. 2011, \apj, 738, 35

\bibitem[{{Kobayashi} {et~al.}(2010){Kobayashi}, {Tanaka}, {Krivov}, \&
  {Inaba}}]{KobayashiEtal2010}
{Kobayashi}, H., {Tanaka}, H., {Krivov}, A.~V., \& {Inaba}, S. 2010, \icarus,
  209, 836

\bibitem[{{Kokubo} \& {Ida}(1998)}]{KokuboIda1998}
{Kokubo}, E. \& {Ida}, S. 1998, Icarus, 131, 171

\bibitem[{{Kokubo} \& {Ida}(2000)}]{KokuboIda2000}
---. 2000, Icarus, 143, 15

\bibitem[{{Lambrechts} \& {Johansen}(2012)}]{LambrechtsJohansen2012}
{Lambrechts}, M. \& {Johansen}, A. 2012, \aap, 544, A32

\bibitem[{{Laughlin} {et~al.}(2004){Laughlin}, {Steinacker}, \&
  {Adams}}]{LaughlinEtal2004i}
{Laughlin}, G., {Steinacker}, A., \& {Adams}, F.~C. 2004, \apj, 608, 489

\bibitem[{{Levison} {et~al.}(2010){Levison}, {Thommes}, \&
  {Duncan}}]{LevisonEtal2010}
{Levison}, H.~F., {Thommes}, E., \& {Duncan}, M.~J. 2010, \aj, 139, 1297

\bibitem[{{Liu} {et~al.}(2012){Liu}, {Madlener}, {Wolf}, {Wang}, \&
  {Ruge}}]{LiuEtal2012}
{Liu}, Y., {Madlener}, D., {Wolf}, S., {Wang}, H., \& {Ruge}, J.~P. 2012, \aap,
  546, A7

\bibitem[{{Markiewicz} {et~al.}(1991){Markiewicz}, {Mizuno}, \&
  {Voelk}}]{MarkiewiczEtal1991}
{Markiewicz}, W.~J., {Mizuno}, H., \& {Voelk}, H.~J. 1991, \aap, 242, 286

\bibitem[{{McCall} {et~al.}(2004){McCall}, {Huneycutt}, {Saykally}, {Djuric},
  {Dunn}, {Semaniak}, {Novotny}, {Al-Khalili}, {Ehlerding}, {Hellberg},
  {Kalhori}, {Neau}, {Thomas}, {Paal}, {{\"O}sterdahl}, \&
  {Larsson}}]{McCallEtal2004}
{McCall}, B.~J., {Huneycutt}, A.~J., {Saykally}, R.~J., {Djuric}, N., {Dunn},
  G.~H., {Semaniak}, J., {Novotny}, O., {Al-Khalili}, A., {Ehlerding}, A.,
  {Hellberg}, F., {Kalhori}, S., {Neau}, A., {Thomas}, R.~D., {Paal}, A.,
  {{\"O}sterdahl}, F., \& {Larsson}, M. 2004, \pra, 70, 052716

\bibitem[{{Meschiari}(2012)}]{Meschiari2012i}
{Meschiari}, S. 2012, \apjl, 761, L7

\bibitem[{{Mizuno}(1980)}]{Mizuno1980}
{Mizuno}, H. 1980, Progress of Theoretical Physics, 64, 544

\bibitem[{{Morbidelli} \& {Nesvorny}(2012)}]{MorbidelliNesvorny2012}
{Morbidelli}, A. \& {Nesvorny}, D. 2012, \aap, 546, A18

\bibitem[{{Mordasini} {et~al.}(2009){Mordasini}, {Alibert}, \&
  {Benz}}]{MordasiniEtal2009}
{Mordasini}, C., {Alibert}, Y., \& {Benz}, W. 2009, \aap, 501, 1139

\bibitem[{{Nelson}(2005)}]{Nelson2005}
{Nelson}, R.~P. 2005, \aap, 443, 1067

\bibitem[{{Nelson} \& {Gressel}(2010)}]{NelsonGressel2010}
{Nelson}, R.~P. \& {Gressel}, O. 2010, \mnras, 409, 639

\bibitem[{{Ogihara} {et~al.}(2007){Ogihara}, {Ida}, \&
  {Morbidelli}}]{OgiharaEtal2007}
{Ogihara}, M., {Ida}, S., \& {Morbidelli}, A. 2007, Icarus, 188, 522

\bibitem[{{Okuzumi}(2009)}]{Okuzumi2009}
{Okuzumi}, S. 2009, \apj, 698, 1122

\bibitem[{{Okuzumi} \& {Hirose}(2011)}]{OkuzumiHirose2011}
{Okuzumi}, S. \& {Hirose}, S. 2011, \apj, 742, 65 (OH11)

\bibitem[{{Okuzumi} \& {Hirose}(2012)}]{OkuzumiHirose2012}
---. 2012, \apjl, 753, L8

\bibitem[{{Okuzumi} \& {Ormel}(2013)}]{OkuzumiOrmel2013}
{Okuzumi}, S. \& {Ormel}, C.~W. 2013, in press (Paper I)

\bibitem[{{Okuzumi} {et~al.}(2009){Okuzumi}, {Tanaka}, \&
  {Sakagami}}]{OkuzumiEtal2009}
{Okuzumi}, S., {Tanaka}, H., \& {Sakagami}, M. 2009, \apj, 707, 1247

\bibitem[{{Ormel} \& {Cuzzi}(2007)}]{OrmelCuzzi2007}
{Ormel}, C.~W. \& {Cuzzi}, J.~N. 2007, \aap, 466, 413

\bibitem[{{Ormel} {et~al.}(2010){Ormel}, {Dullemond}, \&
  {Spaans}}]{OrmelEtal2010i}
{Ormel}, C.~W., {Dullemond}, C.~P., \& {Spaans}, M. 2010, \icarus, 210, 507

\bibitem[{{Ormel} {et~al.}(2012){Ormel}, {Ida}, \& {Tanaka}}]{OrmelEtal2012}
{Ormel}, C.~W., {Ida}, S., \& {Tanaka}, H. 2012, \apj, 758, 80

\bibitem[{{Ormel} \& {Klahr}(2010)}]{OrmelKlahr2010}
{Ormel}, C.~W. \& {Klahr}, H.~H. 2010, \aap, 520, A43

\bibitem[{{Ormel} \& {Kobayashi}(2012)}]{OrmelKobayashi2012}
{Ormel}, C.~W. \& {Kobayashi}, H. 2012, \apj, 747, 115

\bibitem[{{Paardekooper} {et~al.}(2012){Paardekooper}, {Leinhardt},
  {Th{\'e}bault}, \& {Baruteau}}]{PaardekooperEtal2012}
{Paardekooper}, S.-J., {Leinhardt}, Z.~M., {Th{\'e}bault}, P., \& {Baruteau},
  C. 2012, \apjl, 754, L16

\bibitem[{{Paardekooper} \& {Mellema}(2006)}]{PaardekooperMellema2006}
{Paardekooper}, S.-J. \& {Mellema}, G. 2006, \aap, 459, L17

\bibitem[{{Pan} \& {Padoan}(2010)}]{PanPadoan2010}
{Pan}, L. \& {Padoan}, P. 2010, Journal of Fluid Mechanics, 661, 73

\bibitem[{{Perets} \& {Murray-Clay}(2011)}]{PeretsMurray-Clay2011}
{Perets}, H.~B. \& {Murray-Clay}, R.~A. 2011, \apj, 733, 56

\bibitem[{{Pollack} {et~al.}(1996){Pollack}, {Hubickyj}, {Bodenheimer},
  {Lissauer}, {Podolak}, \& {Greenzweig}}]{PollackEtal1996}
{Pollack}, J.~B., {Hubickyj}, O., {Bodenheimer}, P., {Lissauer}, J.~J.,
  {Podolak}, M., \& {Greenzweig}, Y. 1996, Icarus, 124, 62

\bibitem[{{Rafikov}(2006)}]{Rafikov2006}
{Rafikov}, R.~R. 2006, \apj, 648, 666

\bibitem[{{Safronov}(1969)}]{Safronov1969}
{Safronov}, V.~S. 1969, {Evolution of the Protoplanetary Cloud and Formation of
  Earth and the Planets}, ed. V.~S. Safronov (Moscow: Nauka. Transl. 1972 NASA
  Tech. F-677)

\bibitem[{{Sano} {et~al.}(2000){Sano}, {Miyama}, {Umebayashi}, \&
  {Nakano}}]{SanoEtal2000}
{Sano}, T., {Miyama}, S.~M., {Umebayashi}, T., \& {Nakano}, T. 2000, \apj, 543,
  486

\bibitem[{{Shi} \& {Chiang}(2013)}]{ShiChiang2013}
{Shi}, J.-M. \& {Chiang}, E. 2013, \apj, 764, 20

\bibitem[{{Stewart} \& {Leinhardt}(2009)}]{StewartLeinhardt2009}
{Stewart}, S.~T. \& {Leinhardt}, Z.~M. 2009, \apjl, 691, L133

\bibitem[{{Suzuki} {et~al.}(2010){Suzuki}, {Muto}, \&
  {Inutsuka}}]{SuzukiEtal2010}
{Suzuki}, T.~K., {Muto}, T., \& {Inutsuka}, S.-i. 2010, \apj, 718, 1289

\bibitem[{{Tanaka} \& {Ward}(2004)}]{TanakaWard2004}
{Tanaka}, H. \& {Ward}, W.~R. 2004, \apj, 602, 388

\bibitem[{{Teiser} \& {Wurm}(2009)}]{TeiserWurm2009}
{Teiser}, J. \& {Wurm}, G. 2009, \mnras, 393, 1584

\bibitem[{{Thommes} {et~al.}(2003){Thommes}, {Duncan}, \&
  {Levison}}]{ThommesEtal2003}
{Thommes}, E.~W., {Duncan}, M.~J., \& {Levison}, H.~F. 2003, Icarus, 161, 431

\bibitem[{{Thommes} {et~al.}(2008){Thommes}, {Matsumura}, \&
  {Rasio}}]{ThommesEtal2008}
{Thommes}, E.~W., {Matsumura}, S., \& {Rasio}, F.~A. 2008, Science, 321, 814

\bibitem[{{Turner} \& {Drake}(2009)}]{TurnerDrake2009}
{Turner}, N.~J. \& {Drake}, J.~F. 2009, \apj, 703, 2152

\bibitem[{{V\"olk} {et~al.}(1980){V\"olk}, {Jones}, {Morfill}, \&
  {Roeser}}]{VoelkEtal1980}
{V\"olk}, H.~J., {Jones}, F.~C., {Morfill}, G.~E., \& {Roeser}, S. 1980, \aap,
  85, 316

\bibitem[{{Wada} {et~al.}(2009){Wada}, {Tanaka}, {Suyama}, {Kimura}, \&
  {Yamamoto}}]{WadaEtal2009}
{Wada}, K., {Tanaka}, H., {Suyama}, T., {Kimura}, H., \& {Yamamoto}, T. 2009,
  \apj, 702, 1490

\bibitem[{{Weidenschilling}(1977)}]{Weidenschilling1977i}
{Weidenschilling}, S.~J. 1977, \apss, 51, 153

\bibitem[{{Weidenschilling} {et~al.}(1997){Weidenschilling}, {Spaute}, {Davis},
  {Marzari}, \& {Ohtsuki}}]{WeidenschillingEtal1997}
{Weidenschilling}, S.~J., {Spaute}, D., {Davis}, D.~R., {Marzari}, F., \&
  {Ohtsuki}, K. 1997, Icarus, 128, 429

\bibitem[{{Wetherill} \& {Stewart}(1989)}]{WetherillStewart1989}
{Wetherill}, G.~W. \& {Stewart}, G.~R. 1989, Icarus, 77, 330

\bibitem[{{Whipple}(1972)}]{Whipple1972}
{Whipple}, F.~L. 1972, in From Plasma to Planet, ed. A.~{Elvius}, 211

\bibitem[{{Windmark} {et~al.}(2012{\natexlab{a}}){Windmark}, {Birnstiel},
  {G{\"u}ttler}, {Blum}, {Dullemond}, \& {Henning}}]{WindmarkEtal2012}
{Windmark}, F., {Birnstiel}, T., {G{\"u}ttler}, C., {Blum}, J., {Dullemond},
  C.~P., \& {Henning}, T. 2012{\natexlab{a}}, \aap, 540, A73

\bibitem[{{Windmark} {et~al.}(2012{\natexlab{b}}){Windmark}, {Birnstiel},
  {Ormel}, \& {Dullemond}}]{WindmarkEtal2012i}
{Windmark}, F., {Birnstiel}, T., {Ormel}, C.~W., \& {Dullemond}, C.~P.
  2012{\natexlab{b}}, \aap, 544, L16

\bibitem[{{Xie} {et~al.}(2010){Xie}, {Payne}, {Th{\'e}bault}, {Zhou}, \&
  {Ge}}]{XieEtal2010}
{Xie}, J.-W., {Payne}, M.~J., {Th{\'e}bault}, P., {Zhou}, J.-L., \& {Ge}, J.
  2010, \apj, 724, 1153

\bibitem[{{Yang} {et~al.}(2012){Yang}, {Mac Low}, \& {Menou}}]{YangEtal2012}
{Yang}, C.-C., {Mac Low}, M.-M., \& {Menou}, K. 2012, \apj, 748, 79

\bibitem[{{Youdin} \& {Goodman}(2005)}]{YoudinGoodman2005}
{Youdin}, A.~N. \& {Goodman}, J. 2005, \apj, 620, 459

\end{thebibliography}
\end{document}